\documentclass[oneside,reqno]{amsart}

\usepackage{amssymb}
\usepackage{overpic}
\usepackage{graphics}
\usepackage{epsfig}
\usepackage{amsmath}
\usepackage{amsfonts}
\usepackage{paralist}
\usepackage{setspace}
\usepackage{amsthm}
\usepackage{cite}
\usepackage{color}

\newtheorem{theorem}{Theorem}[section]
\newtheorem{corollary}{Corollary}

\newtheorem{proposition}[theorem]{Proposition}

\newtheorem{example}{Example}

\theoremstyle{definition}
\newtheorem{definition}[theorem]{Definition}

\begin{document}

\title[Self-Limiting Trajectories of a Particle]{Self-Limiting Trajectories of a Particle Moving Deterministically in a Random Medium}

\author{B. Z. Webb$^1$}

\author{E. G. D. Cohen$^2$}


\maketitle

\begin{center}
$^{1}$ Brigham Young University, Department of Mathematics, 308 TMCB, Provo, UT 84602, USA\\
$^{2}$ Rockefeller University, Laboratory of Statistical Physics, 1230 York Avenue, New York, NY 10065, USA\\
E-mail: $^1$bwebb@mathematics.byu.edu and $^2$egdc@rockefeller.edu
\end{center}


\begin{spacing}{1}

\begin{abstract}
We study the motion of a particle moving on a two-dimensional honeycomb lattice, whose sites are randomly occupied by either right or left rotators, which rotate the particle's velocity to its right or left, according to deterministic rules. In the model we consider, the scatterers are each initially oriented to the right with probability $p\in[0,1]$. This is done independently, so that the initial configuration of scatterers, which forms the medium through which the particle moves, are both independent and identically distributed. For $p\in(0,1)$, we show that as the particle moves through the lattice, it creates a number of reflecting structures. These structures ultimately \emph{limit} the particle's motion, causing it to have a periodic trajectory. As $p$ approaches either 0 or 1, and the medium becomes increasingly homogenous, the particle's dynamics undergoes a discontinuous transition from this self-limiting, periodic motion to a self-avoiding motion, where the particle's trajectory, away from its initial position, is a self-avoiding walk. Additionally, we show that the periodic dynamics observed for $p\in(0,1)$ can persist, even if the initial configuration of scatterers are not identically distributed. Furthermore, we show that if these orientations are not chosen independently, this can drastically change the particle's motion causing it to have a behavior that is nonperiodic.
\end{abstract}


\section{Introduction}
The particular system we consider in this paper is an example of a Lorentz lattice gas (LLG). In a LLG, a single particle moves along the bonds of a lattice, from lattice site to lattice site. When the particle arrives at a lattice site, it encounters a scatterer that modifies its motion according to a given scattering rule.

The reason to study such systems is to understand the basic principles that underly dynamic processes such as diffusion, propagation, etc. \cite{Ruijgrok88,Wang94,Wang95.3,Bunimovich91,Kong89,Kong90.2,Wang95.1,Meng94,Kong90.1}. For simplicity, the study of a particle's motion on a lattice is a natural choice, since a lattice has both a discrete structure and a high degree of regularity. In such systems the type of scattering rules that have been investigated are physically motivated rules such as rotators, mirrors, etc. \cite{Ruijgrok88,Wang94,Wang95.3,Bunimovich91,Kong90.2,Wang95.1,Bunimovich93,Cao97,Grosfils99,Wang95.2}.

In the case that there is a scatterer at each lattice site and each scatterer is \emph{fixed}, i.e. is not affected by the particle's motion, the problem of determining the particle's motion through the lattice is related to problems in percolation theory \cite{Bunimovich91,Cao97,Ziff91}. When the scatterers are \emph{not fixed}, as is the case in this paper, and are affected by the particle, the particle's motion is a much more dynamic process and has connections to problems in kinetic theory \cite{Bunimovich2002,Velzen1991}.

In this paper, we continue our investigation, begun in \cite{Webb14}, of the particle's motion on the regular two-dimensional honeycomb lattice, in which the lattice is fully occupied by flipping scatterers. The particular type of scatterers we consider here are flipping rotators, which rotate the particle's velocity either to its left or its right by an angle of $\theta=\pm\pi/3$, depending on whether the scatterer is \emph{oriented} to the left or the right, i.e. is a left or right scatterer, respectively. Furthermore, the scatterers \emph{flip} or change orientation after scattering the particle, flipping either from right to left or from left to right, depending on their original orientation, respectively.

This flipping motion of the scatterers has a number of consequences. Viewing the lattice and its scatterers as a medium through which the particle moves, the fact that the particle can change (flip) the scatterer's orientation implies that there is an interplay or interaction between the particle and the medium. That is, the particle affects the medium, which then in turn, has an effect on the particle's motion. It is worth mentioning that this type of system, which we consider in this paper, is likely one of the simplest systems in which there is an interaction between the particle and medium.

Having fixed a lattice and a scattering rule, it remains to choose an initial configuration of the scatterers, in order to study the properties of the motion of the particle over the lattice. The initial configuration we consider, throughout the majority of this paper, is the configuration in which each scatterer is a right rotator with probability $p\in[0,1]$. The orientation of each scatterer is chosen independent from the others, so that the collection of their initial orientations forms a set of independent identically distributed (i.i.d.) random variables. From a physical point of view, we interpret this as an assumption that the particle moves in a random medium.

Although the initial configuration is random, the particle's motion through the lattice is governed by \emph{deterministic} scattering rules. In this setting, the particle's motion is referred to as a \emph{deterministic walk in a random medium} or environment \cite{Bunimovich04}. That is, the first time the particle visits a lattice site, it is scattered to its right by a right scatterer with probability $p\in[0,1]$. However, on each subsequent visit to this same site, the particle is scattered (deterministically) in the opposite direction, from which it was previously scattered, since the scatterer has flipped its orientation after each visit.

We find that, as long as $p\in(0,1)$, the particle's motion will be periodic (see theorem \ref{thm:2}). Because the random initial configuration in this model initially causes the particle to move randomly, this means that, the deterministic nature of the particle's motion eventually overcomes this randomness. The reason this takes place is that, as the particle moves through the lattice, it creates a number of structures, which we refer to as reflecting structures. The initial effect of a reflecting structure on the particle is to cause the particle to reverse its trajectory back to its initial position. However, a second, and arguably more important effect of these structures, is to block the particle's progress through the lattice. We find that a number of reflecting structures, acting together, can permanently limit the particle's motion, causing its trajectory to become periodic (see theorem \ref{thm:1}). Thus, we refer to this particular type of periodic motion as \emph{self-limiting}.

Reflecting structures and their effect on the particle's motion have been previously observed in both the square \cite{Bunimovich93} and the honeycomb lattice \cite{Wang95.1}. In this paper, we can go further and describe the general topology of these structures. We also introduce two other structures and describe their topology and effect on the particle. These new structures are, respectively, called \emph{semi-reflecting structures} and \emph{reflector transforms}.

We show that as the particle moves through the lattice, it can either create, transform, or annihilate a reflecting structure. In order to become trapped or \emph{limited} by a number of reflecting structures, we show that the particle must go through the process of both creating as well as transforming these structures. However, to avoid becoming trapped, the particle can also annihilate those reflectors it has either created or transformed. In this sense, the particle acts like an \emph{architect} that can build, destroy, and remodel, i.e. \emph{transform}, these structures.

This competition between reflector formation, transformation, and annihilation, occurs for all $p$-values in the set $(0,1)$. However, for $p=0,1$ the situation is very different. Building on recent results \cite{Webb14}, we show that for these two $p$-values, the particle can \emph{never} create a reflecting structure (see corollary \ref{cor:2}). Additionally, we observe that the lack of reflecting structures, causes the particle to have a non-periodic and therefore unbounded trajectory \cite{Webb14}. Hence, as $p$ approaches 0 or 1, i.e. as the random medium becomes increasingly homogenous, the particle undergoes a discontinuous dynamical transition from having a periodic, self-limiting motion\footnote{As is described in \cite{Webb14}, this self-avoiding motion is very different from what is used in most computer simulations, since it is deterministically generated and not an ad hoc modified random-walk.} to having a very different non-periodic behavior.

In the periodic, or self-limiting case, the particle becomes trapped between the reflecting structures it creates. Once the particle creates a reflecting structure, it will reverses its entire trajectory back to its initial position, going only through those lattice sites it has already visited. In contrast, for $p=0,1$, the particle not only appears numerically to have an unbounded trajectory but, between returns to its initial position, the particle never visits any lattice site twice (see theorem \ref{thm:1.1}). This self-avoiding motion is then vastly different from the periodic self-limiting motion observed for $p=(0,1)$.

In what seems paradoxical, in a random medium, i.e. for $p\in(0,1)$, the particle has a motion that is \emph{ordered} in the sense that its trajectory is periodic and therefore regularly repeats itself. On the other hand, when $p=0,1$ and the medium is homogeneous, and therefore much more \emph{ordered}, the particle's motion is much more \emph{irregular} exhibiting features of deterministic chaos and of self-organized criticality \cite{Turcotte1999,Webb14}.

To more fully understand how and why reflecting as well as semi-reflecting structures appear in different random media, i.e. for different $p\in(0,1)$, we will also consider a generalization of our flipping rotator model, in which the scatterer's initial configurations are not identically distributed. In that setting we find, under mild conditions, that the particle will still create a number of reflecting structures, which will cause its motion to be periodic (see corollary \ref{cor:4}).

In contrast to this we show, that if there are correlations between the scatterer's initial orientations, i.e. these orientations are not independent, then this can have a noticeable impact on the particle's motion. Specifically, we show that it is possible to have a random configuration with the same distribution as in our original model in this paper, but with \emph{local} correlations between scatters. Interestingly, these local correlations have a global effect on the particle's dynamics, causing it to shift from being periodic to having an unbounded motion (see \cite{Webb14}). As one might expect, although the particle's motion is based on a deterministic rule, its dynamics is also very much dependent on the specific statistical properties of the model's initial configuration.

The paper is organized as follows. In section \ref{sec:2} we introduce the basic model, we will consider throughout the majority of the paper. In section \ref{sec:3} we describe the type of structures the particle can create and annihilate, as it moves through the lattice. We show how these two processes can interact, in such a way, as to cause the particle to have a periodic motion. Based on these processes, we show in section \ref{sec:4} that, for $p\in(0,1)$, the particle in the model we consider will have a periodic trajectory with probability 1.

In section \ref{sec:5} we describe the particle's trajectory for $p=0,1$, where the medium is initially homogenous and the particle has an unbounded and self-avoiding type of motion, as is considered in \cite{Webb14}. We then investigate the transition between the particle's self-limiting and self-avoiding motion, as $p$ approaches 0 and 1. In section \ref{sec:6} we study the case in which the model's initial configuration of scatterers is non-identically distributed, where we find that, under mild conditions, the particle still has a self-limiting motion. In section \ref{sec:7} we introduce other random configurations, in which the scatterers are not independently chosen, to show the effect of local (short range) correlations between initial orientations. We show this can lead to non-periodic dynamics. Some final comments, including a number of open question, are given in section \ref{sec:8}.

We note that although the main results of this paper are proven rigorously, the paper is written so that it can be followed without the need for the reader to work through the various proofs, since the main ideas and techniques used to prove the results in this paper, are described in words, prior to each result.

\begin{figure}
    \begin{overpic}[scale=.3]{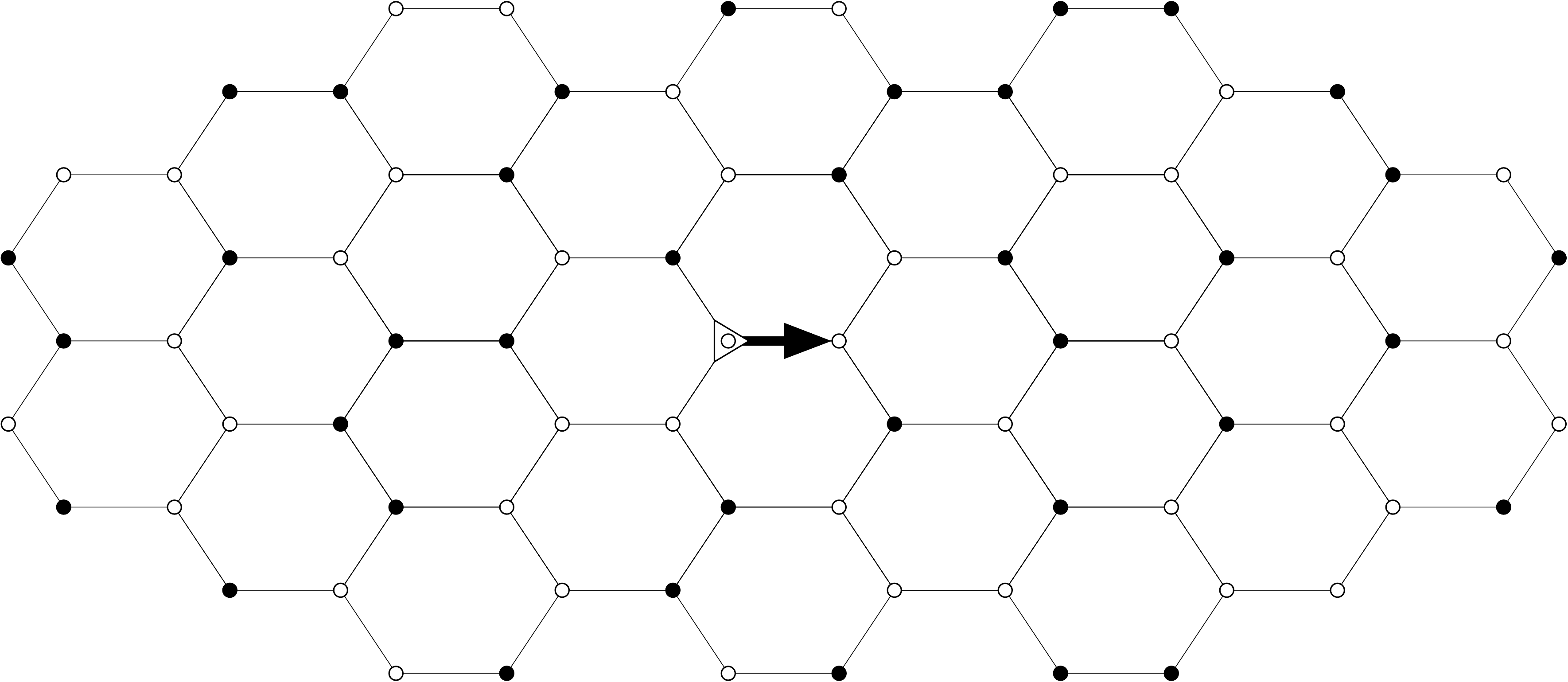}
    \put(43,20.75){$\mathbf{r}$}
    \put(49,24){$\mathbf{v}$}
    \end{overpic}
\caption{A subset of the honeycomb lattice $H=(\mathbb{H},\mathbb{B})$ is shown. The particle's initial position $\mathbf{r}=(0,0)$ and velocity $\mathbf{v}=(1,0)$ are indicated by a triangle and arrow in this paper, respectively. The lattice is shown with a random initial configuration, in which closed circles indicate left scatterers and open circles indicate right scatterers, respectively.}\label{fig:1}
\end{figure}

\section{The Flipping Rotator Model}\label{sec:2}
In this section we describe the particular flipping rotator model, we will consider throughout this paper. The lattice, over which the particle moves, is the honeycomb lattice $H=(\mathbb{H},\mathbb{B})$, with \emph{sites} $\mathbb{H}$ and \emph{bonds} $\mathbb{B}$. This lattice consists of regular hexagons, with sides of unit length, so that each lattice site has three nearest neighbors with which it shares a lattice bond of length 1 (see figure \ref{fig:1}).

Our main object of study in this paper is the motion of a single particle, as it moves from an initial position on the lattice, over the lattice, along the bonds of $H$, from lattice site to lattice site. By way of notation, we let $\mathbf{r}(t)\in\mathbb{R}^2$ denote the \emph{position} and $\mathbf{v}(t)\in\mathbb{R}^2$ denote the \emph{velocity} of the particle at time $t\geq 0$, where the particle is assumed to move with constant unit speed. Moreover, we let $\mathbf{r}=\mathbf{r}(0)$ and $\mathbf{v}=\mathbf{v}(0)$ denote the particle's \emph{initial position} and \emph{initial velocity}, respectively.

For simplicity, we also restrict ourselves to \emph{discrete} time steps $t=0,1,2,\dots$, so that the particle is at some lattice site at each time $t$. The particle's \emph{trajectory} is then the sequence of positions $\{\mathbf{r}(t)\}_{t\geq 0}\subseteq\mathbb{H}$. Since the velocity of the particle does not exist at the moment it is scattered, we let $\mathbf{v}(t)$ denote the velocity of the particle immediately \emph{after} each time step $t\geq 0$.

\begin{figure}
    \begin{overpic}[scale=.35]{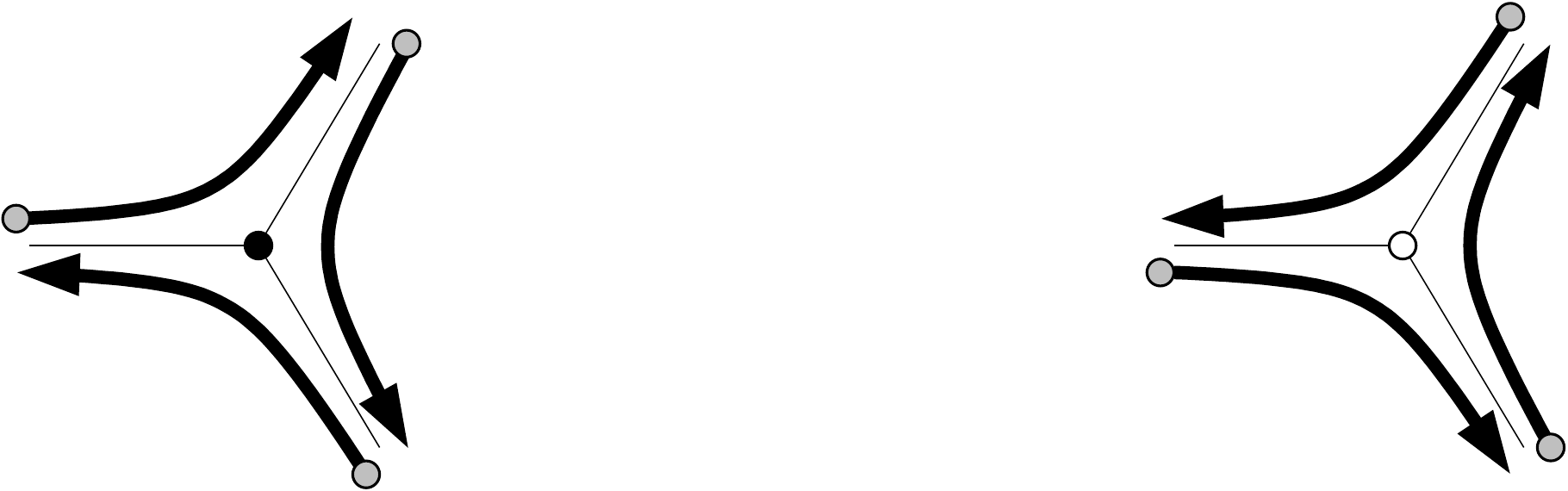}
    \put(-10,-5){(a) \small\emph{left rotators} $(\mathbf{L})$}
    \put(64,-5){(b) \small\emph{right rotators} $(\mathbf{R})$}
    \end{overpic}
    \vspace{0.2in}
\caption{Upon arriving at a left (right) rotator, indicated by a closed (open) circle, the particle's velocity is rotated to its left (right) by an angle of $\theta=\pm\pi/3$. The shaded circles indicate the particle and the arrow its position before and after being scattered.}\label{fig:2}
\end{figure}

At each lattice site $\mathbf{h}\in\mathbb{H}$, we assume that there is a scatterer, which rotates the velocity of the incoming particle, either to its left or to its right, by an angle of $\theta=\pm\pi/3$, depending on the scatterer's orientation. This is shown in figure \ref{fig:2}, where we use the convention, here and throughout the paper, that a closed circle denotes a left rotator and an open circle denotes a right rotator, respectively. The assumption, that there is a scatterer at each lattice site of $H$, ensures that the particle remains on the lattice for all $t\geq0$.

Note that each scatterer is, initially, either a left scatterer or a right scatterer, respectively, i.e. is \emph{oriented} either to the left or to the right. With this in mind, we let $C=C(0)$ denote the \emph{initial configuration} of scatterers and let $C(t)$ denote the \emph{configuration} of scatterers on the lattice at time $t\geq 0$. The configuration $C(t)$ for $t\geq 0$ consists of the collection of all the individual orientations of each scatter on the lattice at time $t$. For each lattice site $\mathbf{h}\in\mathbb{H}$ we let
\begin{equation*}
C(t,\mathbf{h})\in\{-1,1\} \ \ \text{for} \ \mathbf{h}\in\mathbb{H}, \  t\geq 0,
\end{equation*}
denote the \emph{orientation} of the scatterer at the site $\mathbf{h}$ at time $t$. The orientation $C(t,\mathbf{h})=-1$ indicates, that at time $t$ the scatterer at lattice site $\mathbf{h}$ is a left scatterer, whereas the orientation $C(t,\mathbf{h})=1$ indicates, that the scatterer is a right scatterer. Furthermore, we let $C(\mathbf{h})\equiv C(0,\mathbf{h})$ denote the \emph{initial orientation} of the scatterer at $\mathbf{h}\in\mathbb{H}$ at time $t=0$.

Suppose the particle has initial position $\mathbf{r}$ and initial velocity $\mathbf{v}$. For an initial configuration $C$, we call $I=(\mathbf{r},\mathbf{v},C)$ an \emph{initial condition}. For an initial condition $I$, the particle's \emph{deterministic} equations of motion are given by
\begin{align}
\mathbf{r}(t+1)&=\mathbf{r}(t)+\mathbf{v}(t), \label{eq:1}\\
\mathbf{v}(t+1)&=R\big[C(t,\mathbf{r}(t+1))\big]\mathbf{v}(t), \label{eq:2}\\
C(t+1,\mathbf{h})&=
\begin{cases}
-C(t,\mathbf{h}) &  \ \ \text{if} \ \  \mathbf{h}=\mathbf{r}(t+1)\\
\hspace{0.1in} C(t,\mathbf{h}) & \ \ \text{otherwise}\label{eq:3}
\end{cases}
\end{align}
for $t\geq 0$. Equation \eqref{eq:1} gives the dynamics of the particle, describing its piecewise linear motion between successive scatterings. The \emph{rotation operator} $R:\{-1,1\}\rightarrow\mathbb{R}^{2\times 2}$ in equation (\ref{eq:2}), is the rotation matrix given by
\begin{equation}\label{eq:matrix}
R[z]=
\left[\begin{array}{rr}
\cos(\frac{\pi}{3}z)&\sin(\frac{\pi}{3}z)\\
-\sin(\frac{\pi}{3}z)&\cos(\frac{\pi}{3}z)
\end{array}\right] \ \text{where} \ z\in\{-1,1\},
\end{equation}
which describes how the particle's velocity is rotated, when it arrives at a scatterer. Equation \eqref{eq:3} describes the flipping motion of the scatterers.

Given an initial condition $I$, the particle's motion over the lattice is uniquely determined for all $t\geq0$ by equations \eqref{eq:1}--\eqref{eq:3}. This leads us to the following definition, which describes the general type of LLG we consider in this paper.

\begin{definition}\label{def:frm}
Let $(H,I)$ denote the LLG with the initial condition $I=(\mathbf{r},\mathbf{v},C)$ and equations of motion given by the equations \eqref{eq:1}--\eqref{eq:3}. We will call this the \emph{flipping rotator model} on the honeycomb lattice with the initial condition $I$.
\end{definition}

In general, one can consider two types of initial configurations. The first are \emph{fixed} or \emph{deterministic initial configurations}, which are specific configurations typically given by some deterministic rule. For example, the initial configuration $C$ in definition \ref{def:frm}, although arbitrary, is a deterministic configuration, since we assume that it has been specified. These deterministic type of configurations are considered throughout \cite{Webb14}.

The second type of initial configurations we consider are \emph{random initial configurations}, which are generated according to some probabilistic rule. A major difference between a deterministic configuration of scatterers and one that is randomly generated is that, a deterministic configuration is a single unique configuration, while a randomly generated configuration is a \emph{realization} of a random process, of which there are typically many realizations.

In this paper, our main focus is on random initial configurations. The particular type of initial configurations we consider here, will be generated based on the following probabilistic rule. Let $(H;p)$ denote the flipping rotator model, in which each scatterer is initially a right scatterer with probability $p$, for $p\in[0,1]$. We furthermore assume that, the initial orientation of one scatterer does not influence the initial orientation of any other. Hence, the collection of initial orientations $C$ in the $(H;p)$ model forms a collection of \emph{independent identically distributed} (i.i.d.) random variables, in which the probability
\begin{equation}\label{assump1}
\mathbb{P}[C(\mathbf{h})=1]=p \ \text{for all} \ \mathbf{h}\in\mathbb{H}.
\end{equation}
Since this model has a random initial configuration we may assume, without loss of generality, that the particle's initial position and initial velocity are
\begin{equation}\label{assump2}
\mathbf{r}=(0,0) \ \text{and} \ \mathbf{v}=(1,0),
\end{equation}
respectively, as is shown in figure \ref{fig:1}. That is, the particle in the $(H;p)$ model is at the origin of the lattice at time zero and moving to the right.

Under these assumptions we define the following model, which is similar to the flipping rotator model described in definition \ref{def:frm}. The main difference is that its initial configuration is randomly generated.

\begin{definition}\label{def:frmprob}
For $p\in[0,1]$, let $(H,p)$ denote the LLG with the random initial condition $I=(\mathbf{r},\mathbf{v},C)$ given by equations \eqref{assump1}--\eqref{assump2}, and equations of motion given by equations \eqref{eq:1}--\eqref{eq:3}. We call this the \emph{flipping rotator model} on the honeycomb lattice with the \emph{random initial condition} $I$.
\end{definition}

Although the particle's dynamics are deterministic in the $(H;p)$ model, the model's initial configuration is random for each $p\in(0,1)$. Therefore, the particle's motion is what is referred to as a deterministic walk in a random environment \cite{Bunimovich04} or, alternatively, in a random medium. Here, we refer to this type of motion as a \emph{pseudo-random walk}, since it has aspects of both random and deterministic motion. An alternative, but equivalent, way of describing the particle's motion in the $(H;p)$ model is to say that, the particle's trajectory is a random walk up to the point at which it returns to a lattice site it has already visited. Once the particle returns to a site it has visited, its motion is determined by equations \eqref{eq:1}--\eqref{eq:3}.

Despite the randomness introduced by the initial configuration in the $(H;p)$ model, we find the particle's trajectory, over the lattice, is qualitatively the same for any realization of this initial configuration. Specifically, for each $p\in(0,1)$, the particle's trajectory in the $(H;p)$ model will always be \emph{periodic}. That is, although the random medium, through which the particle moves, will initially cause the particle to have a random-like motion, ultimately the particle's deterministic behavior will overcome this randomness and its motion will be periodic.

By periodic we mean that, there is a $t_p<\infty$, such that the particle's position satisfies $\mathbf{r}(t)=\mathbf{r}(t+t_p)$ for each $t\geq 0$, where $t_p$ is the particle's \emph{period}. The fact that the particle's motion in the $(H;p)$ model is always periodic for $p\in(0,1)$, is one of the main results of this paper, which can be stated as follows.

\begin{theorem}\label{thm:2} \textbf{(Self-Limiting Trajectories)}
For $p\in(0,1)$, the particle's trajectory in the $(H;p)$ model will be periodic with probability 1.
\end{theorem}

The proof of theorem \ref{thm:2} is postponed until section \ref{sec:4}. The reason for this is, that we first need an understanding of the kinds of ``structures" the particle can create, as it moves through the lattice and how these structures ultimately \emph{limit} the particle's motion. Here, the term \emph{structure} refers to a collection of scatterers, which \emph{collectively} have a specific effect on the particle's motion. The type of structure, which is most important to the particle's dynamics in this model, is what we will refer to as a \emph{reflecting structure}.

\begin{figure}
\begin{center}
\begin{tabular}{ccc}
    \begin{overpic}[scale=.56]{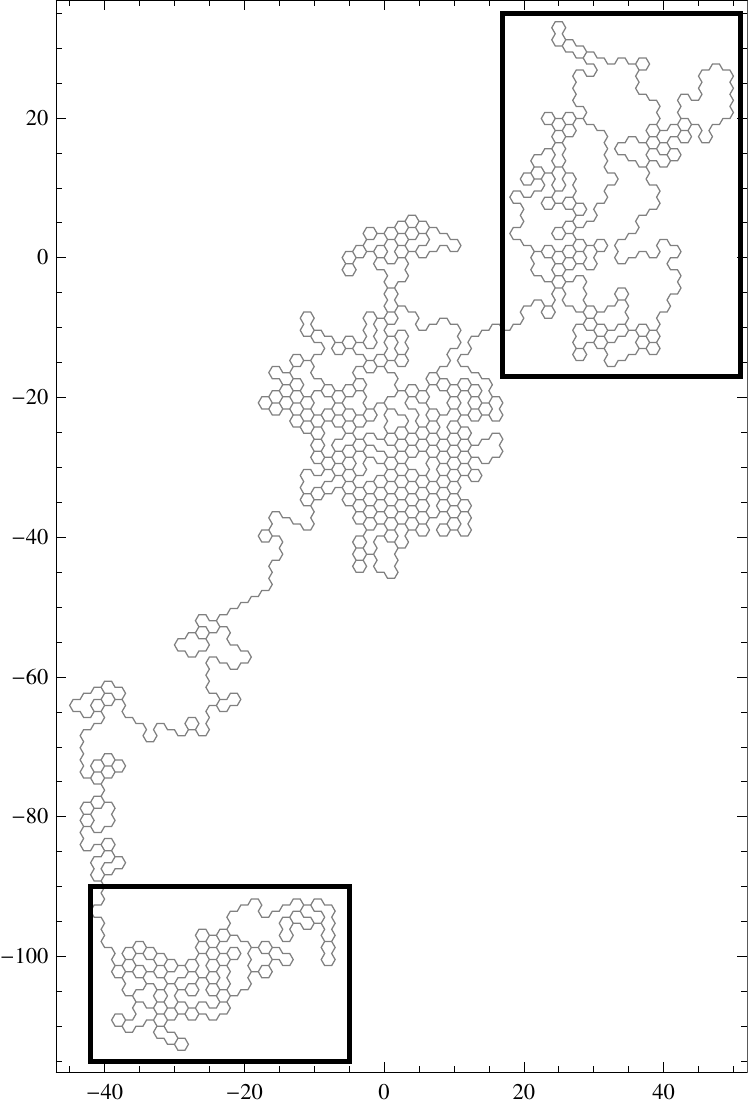}
    \put(17,21.5){$\Psi_2$}
    \put(54,60){$\Psi_1$}
    \put(29.5,-5){$p=0.5$}
    \end{overpic} &
    \begin{overpic}[scale=.415]{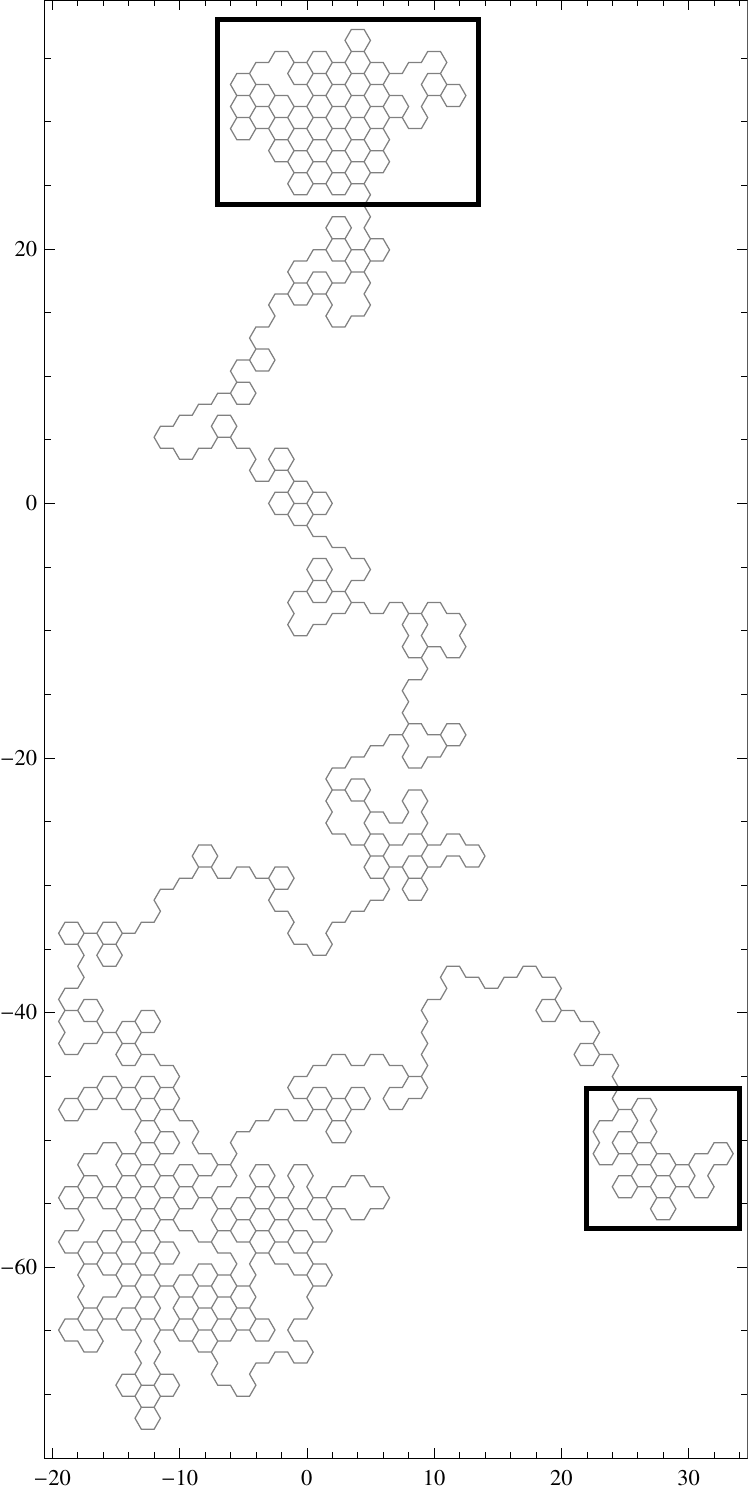}
    \put(42,12){$\Psi_2$}
    \put(34,90){$\Psi_1$}
    \put(19.5,-5){$p=0.7$}
    \end{overpic} &
    \begin{overpic}[scale=.5025]{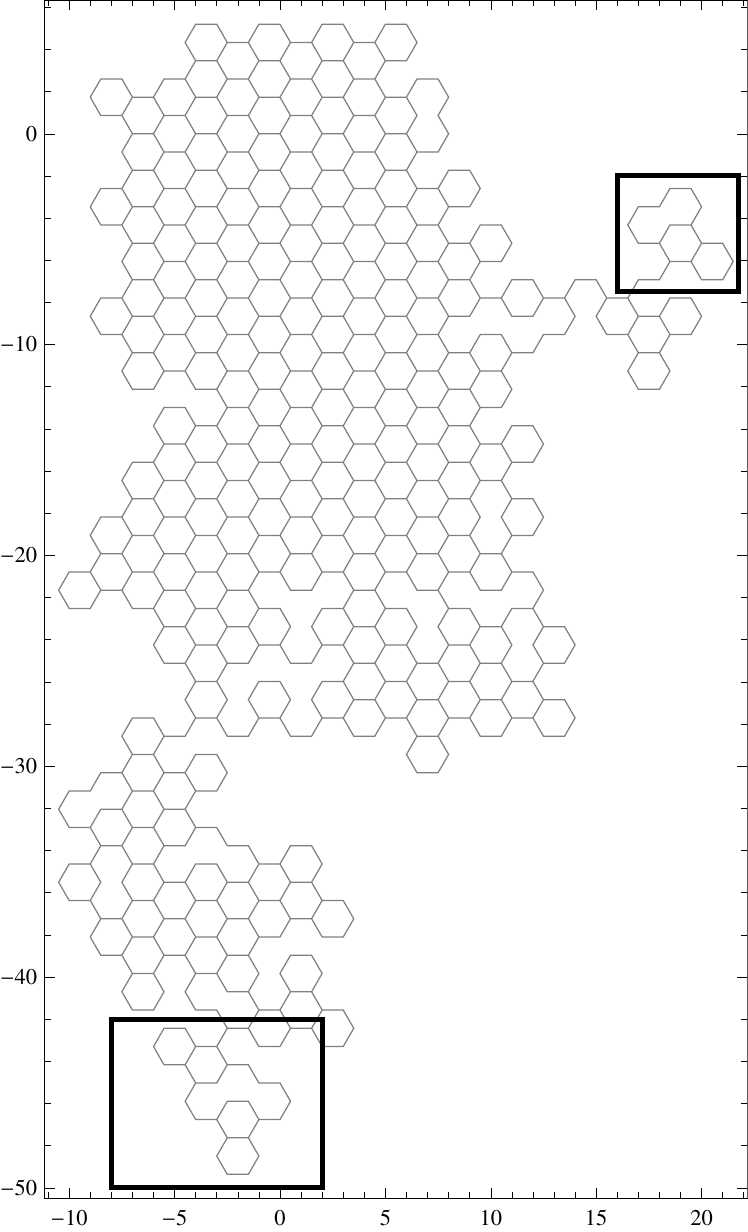}
    \put(27.5,8){$\Psi_2$}
    \put(52.5,87.5){$\Psi_1$}
    \put(20.5,-5){$p=0.9$}
    \end{overpic}
\end{tabular}
\end{center}
\caption{A realization of the particle's trajectory in the $(H;p)$ model for $p=0.5,0.7$, and $0.9$. In each case the particle's trajectory is periodic, which is due to the formation of two different reflecting structures $\Psi_1$ and $\Psi_2$, between which the particle's entire trajectory oscillates. The periods of these trajectories are respectively $t_p=11516,7944$, and $22504$.}\label{fig:3.0}
\end{figure}
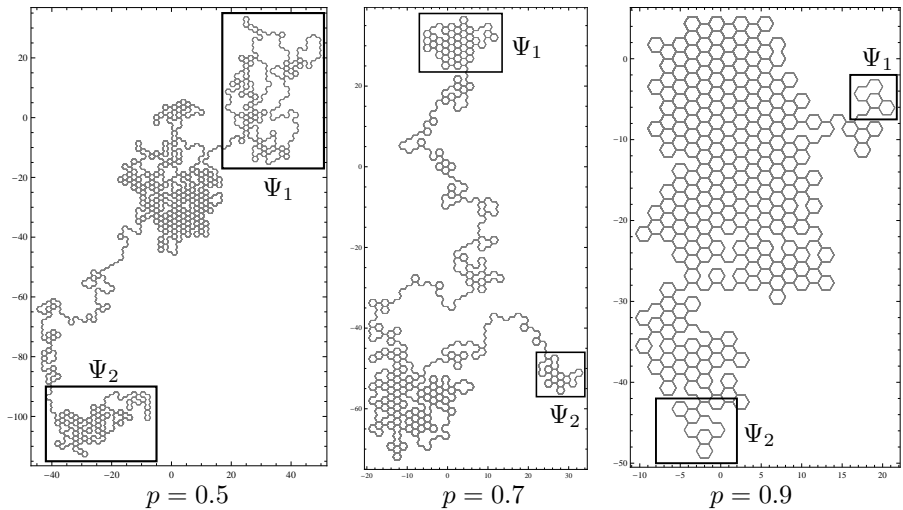

A reflecting structure is, in effect, a collection of scatterers that causes the particle to reverse its trajectory back to its initial position (see definition \ref{def:refstruct}). A number of reflecting structures can be seen in figure \ref{fig:3.0}, where a particular realization of the particle's trajectory in the $(H;p)$ model is shown for $p=0.5,0.7,0.9$, respectively. In each case there are two reflecting structures $\Psi_1$ and $\Psi_2$ which, acting together, cause the particle to have a periodic trajectory and thereby ``limit" the particle's motion to a finite subset of the lattice (see theorem \ref{thm:1}).

In the following section we will show that, as the particle moves through the lattice it, acting as an ``architect" of sorts, can either \emph{create}, \emph{transform}, or \emph{annihilate} a reflecting structure. Moreover, each one of these processes is an integral part of particle's dynamics in the $(H;p)$ model and each one will be used to prove theorem \ref{thm:2}.

Before continuing, it should be pointed out that in other flipping rotator models, where the particle moves on some lattice other than the honeycomb lattice, the particle will have a different motion, than that described in theorem \ref{thm:2}. For instance, on the triangular lattice it has been shown that, for any initial configuration of scatterers on the lattice, the particle will always propagate in a strip away from its initial position \cite{Grosfils99}. On the square lattice, in which every lattice site is occupied by a scatterer, it has been shown that the particle will have an unbounded non-periodic trajectory irrespective of the initial configuration of scatterers \cite{Bunimovich91}.

Before finishing this section, it should be noted that the $(H;p)$ model does not have a random initial configuration for all $p$-values. In particular, for $p=0,1$ the initial configuration is the deterministic configuration consisting of all left and all right scatterers, respectively, considered in the previous paper \cite{Webb14}. These configurations will be important in this paper and are discussed later in section \ref{sec:5}.

Recall that one can think of the lattice, together with its scatterers, as a medium through which the particle moves. In this context, $p$ can be used as a measure of how homogenous or inhomogeneous this medium is. With this in mind, one of our goals in this paper is to understand the difference between the particle's dynamics when $p\in(0,1)$ and when $p=0,1$. That is, we are investigating how the degree of homogeneity (or lack thereof), in a random lattice medium, affects the particle's dynamics. In particular, we are interested in the dynamical transition the particle undergoes, as the lattice medium becomes more homogenous, i.e. as $p$ approaches 0 and 1.

\section{Creation, Transformation, and Annihilation of Reflecting Structures}\label{sec:3}
Before studying the $(H;p)$ model, we first consider the $(H,I)$ model in which $I$ is an arbitrary but fixed initial condition. The reason we do this is that, to fully understand the dynamics in the $(H;p)$ model, we need first, to describe the type of structures that the particle can form in the $(H,I)$ model.

There are, in fact, two types of structures we will consider in this section. These are reflecting and semi-reflecting structures, respectively. \emph{Reflecting structures} cause the particle to return to its initial position via the same sequence of positions it took to reach the reflector, only in reverse. In contrast, a \emph{semi-reflecting structure} may reverse the particle's trajectory only part of the way back to its initial position, at which point the particle begins moving along a different sequence of lattice sites.

We begin by describing how the particle is able to \emph{create} these structures. We then show how the particle can, first, transform one reflector into another reflector and, second, how the particle can annihilate a reflector. In the following section, section \ref{sec:4}, the interplay between reflector creation, transformation, and annihilation will be used to prove theorem \ref{thm:2} (see section \ref{sec:2}).

To describe the structure of a reflector and semi-reflector, we need the following definitions. For the LLG $(H,I)$, let $T[t_1,t_2]=\{\mathbf{r}(t):t_1\leq t\leq t_2\}$, so that $T[t_1,t_2]$ denotes the \emph{trajectory} of the particle from time $t_1$ to time $t_2$. The particle in $(H,I)$ is said to travel on a \emph{loop}
$L=T[t_1,t_2]$ from time $t_1$ to time $t_2$, if $\mathbf{r}(t_1)=\mathbf{r}(t_2)$ and $\mathbf{r}(t)\neq\mathbf{r}(t_1)$ for all $t_1<t<t_2$. In this case we call the position $\mathbf{r}(t_1)=\mathbf{r}(t_2)$ the \emph{base} of $L$.

The reflecting structures, that will be important for understanding the particle's dynamics, are defined as follows.

\begin{definition}\label{def:refstruct}
Suppose there are times $0<t_1<t_*<t_2$ in the LLG $(H,I)$, where\\
(a) $L_1=T[t_1,t_*]$ and $L_2=T[t_*,t_2]$ are loops;\\
(b) both $L_1$ and $L_2$ intersect $\mathbf{r}(t_1+1)$ and $\mathbf{r}(t_2-1)$ exactly once; and\\
(c) $T[0,t_1-1]\cap T[t_1,t_2]=\emptyset$.\\
Then we call the particle's sequence of positions $\Psi=T[t_1,t_2]$ a \emph{reflecting structure} of $(H,I)$, \emph{based} at the point $\mathbf{r}(t_1)=\mathbf{r}(t_2)$. We say that the particle \emph{encounters} the reflecting structure $\Psi=T[t_1,t_2]$ at time $t_1$.
\end{definition}

\begin{figure}
    \begin{overpic}[scale=.29]{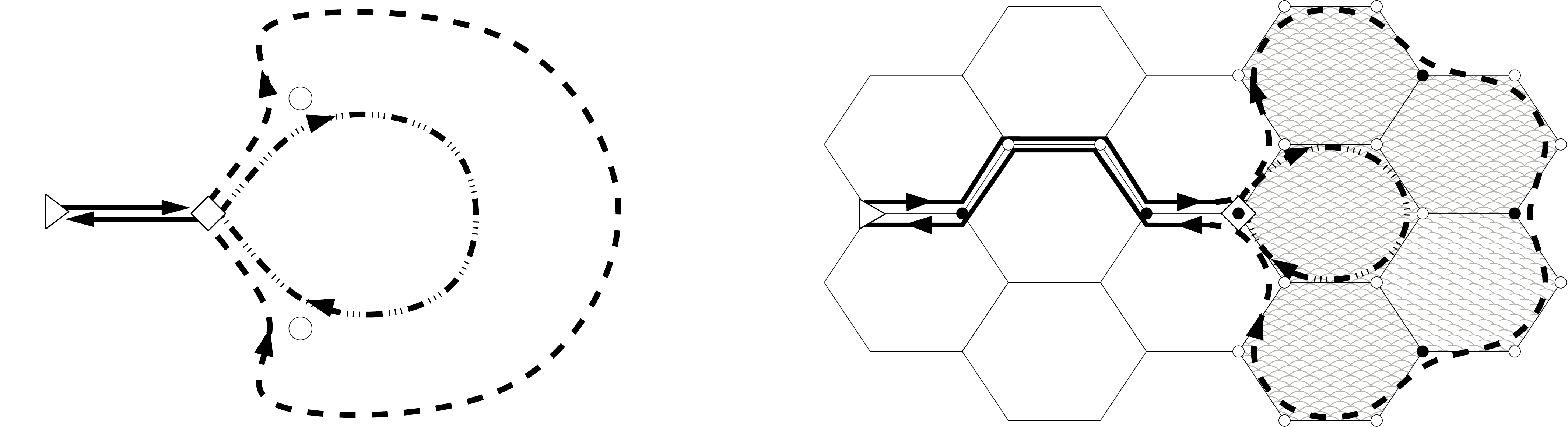}
    \put(15.5,13.25){\small$\mathbf{r}(t_1)=\mathbf{r}(t_2)$}
    \put(20,22.5){\small$\mathbf{r}(t_1+1)$}
    \put(20,3.5){\small$\mathbf{r}(t_2-1)$}
    \put(1,13.15){\small$\mathbf{r}$}
    \put(53,13.15){\small$\mathbf{r}$}
    \put(85,12.75){$\Psi$}
    \put(8,-3){\emph{(a) reflector topology}}
    \put(53,-3){\emph{(b) example of a reflecting structure}}
    \end{overpic}
    \vspace{0.1in}
\caption{The general topology of a reflecting structure is shown in (a). An example of a reflecting structure $\Psi$ and the particle's return to the origin is shown in (b). In both cases, the base of the reflecting structure, at $\mathbf{r}(t_1)=\mathbf{r}(t_2)$, is indicated by a diamond. The reflector's two loops are shown as dashed-dotted and dashed lines, respectively. Solid lines indicate the particle's trajectory before and after the reflecting structure, where the particle moves first away from and then returns to its initial position $\mathbf{r}$.}\label{fig:3}
\end{figure}
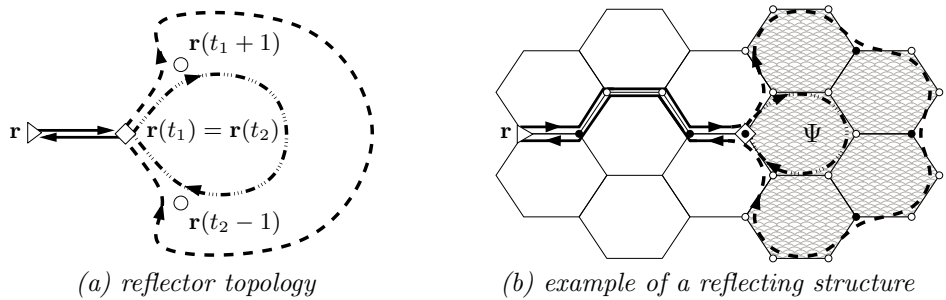

From definition \ref{def:refstruct}, the reflecting structure $\Psi$ consists of two loops, which are both based at the same point $\mathbf{r}(t_1)=\mathbf{r}(t_2)$. This point, which is also the \emph{base} of the reflecting structure, is visited exactly three times by the particle as it moves through the reflector (cf. figure \ref{fig:3}).

These loops can have any (finite) shape or size on the lattice, with the following restrictions. First, the loops cannot intersect any of the particle's trajectory, prior to the reflecting structure. Second, each of these loops must visit the two lattice sites $\mathbf{r}(t_1+1)$ and $\mathbf{r}(t_2-1)$ adjacent to the reflecting structure's base exactly once. This topological structure of a reflector is shown in figure \ref{fig:3}(a). An example of a reflecting structure on the honeycomb lattice $H$ is shown in figure \ref{fig:3}(b).

When the particle encounters a reflecting structure, it will move through the structure and exit it at the same point it entered, but in the opposite direction. From there, the particle will retrace its trajectory, prior to the reflecting structure, back to its initial position. In this sense, a reflecting structure acts as a ``mirror", which effectively reflects the particle back along its original trajectory. This physical description of a reflector's effect on the particle summarizes the following proposition.

\begin{proposition}\label{prop:reflect} \textbf{(Reflecting Property)}: If $\Psi=T[t_1,t_2]$ is a reflecting structure of $(H,I)$, then $\mathbf{r}(t_2+t)=\mathbf{r}(t_1-t)$ for all $0\leq t\leq t_1$. In particular, the particle returns to its initial position at time $t=t_2+t_1$.
\end{proposition}

Before proving proposition \ref{prop:reflect}, we need the following notation. If $\Omega\subset H$, let $C_{\Omega}(t)$ denote the configuration of scatterers on the subset $\Omega\subset H$ of the honeycomb lattice at time $t\geq 0$. Moreover, let $C_{\bar{\Omega}}(t)$ denote the configuration of scatterers on the subset $\bar{\Omega}=H-\Omega$, which is the \emph{complement} of $\Omega$.

The proof of this proposition is based on the following observation. Suppose the particle is scattered to the right (left) by a scatterer at some lattice site $\mathbf{r}(t)$, at time $t>0$. If immediately after being scattered, the particle were, for some reason, to reverse its velocity and head back towards this scatter, it would find, upon returning, that the scatterer's orientation had flipped from right (left) to left (right). This change in orientation would then cause the particle to be scattered back to the lattice site $\mathbf{r}(t-1)$, it had visited at time $t-1$.

That is, any event that causes the particle to be ``reflected" back towards a lattice site it has previously visited, will cause the particle to reverse its entire trajectory back to its initial position. In this model, a mechanism that can cause such a reflection is a reflecting structure. This observation is the main argument used in the following proof of proposition \ref{prop:reflect}.

\begin{proof}
Suppose $\Psi=T[t_1,t_2]$ is a reflecting structure of $(H,I)$, where $I=(\mathbf{r},\mathbf{v},C)$. Then $\mathbf{r}(t_2)=\mathbf{r}(t_1)$, since the particle is at the base of $\Psi$ at times $t_1$ and $t_2$. Moreover, parts (a) and (b) of definition \ref{def:refstruct} imply that $\mathbf{v}(t_2)=-\mathbf{v}(t_1-1)$ and part (c) implies then that the configuration $C_{\bar{\Psi}}(t_2)=C_{\bar{\Psi}}(t_1-1)$.

Continuing by induction, suppose for some $t\in[0,t_1-1)$, that $\mathbf{r}(t_2+t)=\mathbf{r}(t_1-t)$, $\mathbf{v}(t_2+t)=-\mathbf{v}(t_1-t-1)$, and $C_{\bar{\Psi}}(t_2+t)=C_{\bar{\Psi}}(t_1-t-1)$. Then equation \eqref{eq:1} implies that
\begin{align*}
\mathbf{r}(t_2+t+1)=\mathbf{r}(t_1-t)-\mathbf{v}(t_1-t-1)=\mathbf{r}(t_1-t-1).
\end{align*}
From equation \eqref{eq:3} it then follows that
\begin{align*}
C_{\bar{\Psi}}(t_2+t+1,\mathbf{h})=
\begin{cases}
-C_{\bar{\Psi}}(t_1-t-1,\mathbf{h}) \ \ \text{if} \ \ \mathbf{h}=\mathbf{r}(t_1-t-1)\\
\hspace{0.1in} C_{\bar{\Psi}}(t_1-t-1,\mathbf{h}) \ \ \text{otherwise,}
\end{cases}
\end{align*}
for each $\mathbf{h}\in\bar{\Psi}$. Hence, $C_{\bar{\Psi}}(t_2+t+1)=C_{\bar{\Psi}}(t_1-t)$. Additionally, we have
\begin{align*}
\mathbf{v}(t_2+t+1)=R&\big[C(t_2+t,\mathbf{r}(t_2+t+1))\big] \mathbf{v}(t_2+t)=\\
-R&\big[C(t_1-t-1,\mathbf{r}(t_1-t-1))\big] \mathbf{v}(t_1-t-1)= -\mathbf{v}(t_1-t-2),
\end{align*}
since $R[C(t_1+t-1,\mathbf{r}(t_1+t-1))]=R^{-1}[C(t_1+t-2,\mathbf{r}(t_1+t-1))]$.

By induction, it then follows that $\mathbf{r}(t_2+t)=\mathbf{r}(t_1-t)$, $\mathbf{v}(t_2+t)=-\mathbf{v}(t_1-t-1)$, and $C_{\bar{\Psi}}(t_2+t)=C_{\bar{\Psi}}(t_1-t-1)$, for all $0\leq t\leq t_1-1$. For $t=t_1$ this implies that
\begin{align*}
\mathbf{r}(t_2+t_1)=\mathbf{r}(t_2+t_1-1)+\mathbf{v}(t_2+t_1-1)= \mathbf{r}(1)-\mathbf{v}(0)=\mathbf{r}(0),
\end{align*}
completing the proof.
\end{proof}

Proposition \ref{prop:reflect} states that after the particle exits a reflecting structure, its motion will be on the same sequence of lattice sites, as it took to arrive at the structure, except in the opposite direction. Having established this ``reflection property," we give an example of the creation or \emph{formation} of a reflecting structure.

\begin{example} \textbf{(Reflector Formation)}
Consider the trajectory of the particle shown in figure \ref{fig:3}(b), consisting of the solid, dashed, and dashed-dotted lines. Note that, by definition \ref{def:refstruct}, $\Psi=T[5,27]$ is a reflecting structure so that, as guaranteed by proposition \ref{prop:reflect}, the particle's position $\mathbf{r}(27+t)=\mathbf{r}(5-t)$ for $0\leq t\leq 5$. In particular, the particle returns to its initial position at time $t=32$ by reversing its trajectory back to its initial position, as can be seen in the figure.
\end{example}

There are three reasons why a reflecting structure returns the particle to its initial position, based on the proof of proposition \ref{prop:reflect}. The first is that the particle leaves the reflector from the same point it entered, but in the opposite direction. Next, because of the flipping motion of the scatterers, the particle then begins to retrace each one of its previous steps, back to its initial position. The third, is that $\Psi=T[t_1,t_2]$, as a collection of lattice sites, does not overlap with the particle's initial trajectory $T[0,t_1]$ before time $t_1$. Therefore, the particle's motion through the reflector does not affect the configuration of the scatterers along the particle's return path, back to its initial position.

If a reflecting structure $\Psi=T[t_1,t_2]$ does happen to intersect the particle's trajectory prior to time $t_1$, then the structure is no longer a reflecting structure in the sense of definition \ref{def:refstruct} (see part (c)). This structure is, however, still similar to a reflecting structure, both in terms of its topology and its effect on the particle, which we will describe below. We refer to such structures as \emph{semi-reflecting structures}, which are defined as follows.

\begin{definition}\label{def:semi}
Suppose in the LLG $(H,I)$, that the sequence of positions $\Phi=T[t_1,t_2]$ is a reflecting structure, with the exception that $\mathbf{r}(\tau)\in T[t_1,t_2]$ for some $0<\tau<t_1$. Then we call $\Phi$ a \emph{semi-reflecting structure} of $(H,I)$.
\end{definition}

\begin{figure}
    \begin{overpic}[scale=.295]{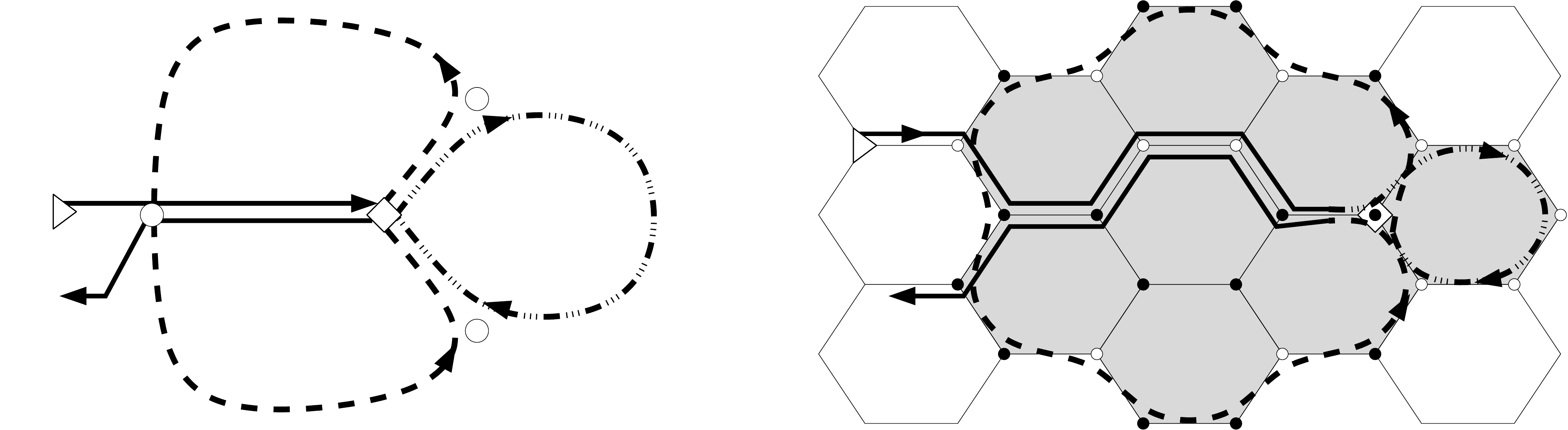}
    \put(26.5,13.25){\small$\mathbf{r}(t_1)=\mathbf{r}(t_2)$}
    \put(31.5,22){\small$\mathbf{r}(t_1+1)$}
    \put(31.5,4.25){\small$\mathbf{r}(t_2-1)$}
    \put(10.5,10.75){\small$\mathbf{r}(\tau)$}
    \put(1.5,13.25){\small$\mathbf{r}$}
    \put(52.25,17.5){\small$\mathbf{r}$}
    \put(75,13.25){$\Phi$}
    \put(5,-3){\emph{(a) semi-reflector topology}}
    \put(46,-3){\emph{(b) an example of a semi-reflecting structure}}
    \put(58,13){\small$\mathbf{r}(\tau)$}
    \end{overpic}
    \vspace{0in}
\caption{The topology of a semi-reflecting structure is shown in (a). An example of a semi-reflecting structure $\Phi$ is shown in (b) and its affect on the particle's trajectory. In both (a) and (b), the base of the semi-reflecting structure at $\mathbf{r}(t_1)=\mathbf{r}(t_2)$ is indicated by a diamond. The semi-reflector's two loops are shown as dashed-dotted and dashed lines respectively. Solid lines indicate the particle's trajectory before and after the reflecting structure, which demonstrate, in both figures (a) and (b), how the particle stops reversing its trajectory at the lattice site $\mathbf{r}(\tau)$.}\label{fig:4}
\end{figure}
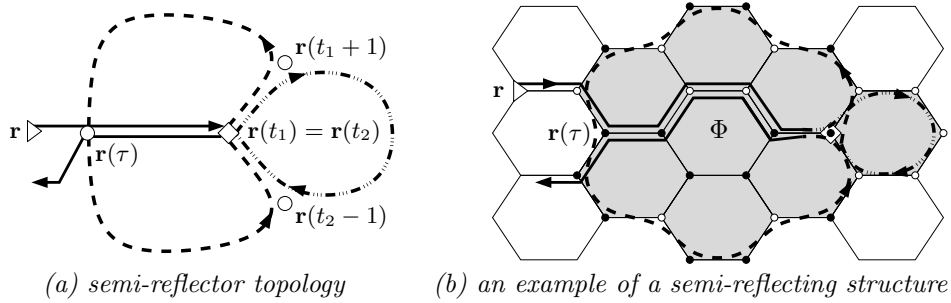

The topology of a general semi-reflecting structure is shown in figure \ref{fig:4}(a) and an example of a semi-reflecting structure is shown in \ref{fig:4}(b).Similar to a reflecting structure, a semi-reflecting structure $\Phi=T[t_1,t_2]$ consists of two loops, both of which are based at the same point $\mathbf{r}(t_1)=\mathbf{r}(t_2)$. Additionally, both loops also visit the sites $\mathbf{r}(t_1+1)$ and $\mathbf{r}(t_2-1)$ adjacent to their base exactly once. The difference, however, between a reflecting structure and a semi-reflecting structure is that, as the particle moves through a semi-reflector it passes through at least one lattice site it has already visited prior to encountering this structure. Therefore, part (c) of definition \ref{def:refstruct} does not hold for semi-reflectors.

The physical effect a semi-reflecting structure has on the particle's motion, is the following. When the particle exits a semi-reflecting structure, it will begin to reverse its trajectory back to its initial position, as if it had encountered a reflecting structure. However, since the semi-reflector intersects at least one of the lattice sites $\mathbf{r}(\tau)$ that the particle has previously visited, the particle, by moving through the semi-reflector, will have altered this scatterer's orientation. The idea is that, when the particle arrives back at the lattice site $\mathbf{r}(\tau)$, the difference in the scatterer's orientation at this site, will cause the particle to deviate from its original trajectory. In this way, the semi-reflecting structure will reverse the particle's trajectory back towards its initial position but, before the particle arrives at its initial position, this process of reversal will stop. This is summarized in the following proposition.

\begin{proposition}\label{prop:semi} \textbf{(Semi-Reflecting Property)}: Let $\Phi=T[t_1,t_2]$ be a semi-reflecting structure of $(H,I)$. If $\tau<t_1$ is the largest time such that $\mathbf{r}(\tau)\in \Phi$, then $\mathbf{r}(t_1-t)=\mathbf{r}(t_2+t)$ for all $0\leq t\leq t_1-\tau$.
\end{proposition}

To prove this proposition we first note that, if the particle encounters a semi-reflecting structure $\Phi$, then this structure does in fact act like a \emph{true} reflecting structure up to the point when the particle returns to the lattice site $\mathbf{r}(\tau)$.  Since $\Phi$ can be considered, up to this point in time, to be a reflecting structure, proposition \ref{prop:reflect} can be used to prove proposition \ref{prop:semi}. This is the main technique we use in the proof of this result.

\begin{proof}
Suppose $\Phi=T[t_1,t_2]$ is a reflecting structure of $(H,I)$ where $I=(\mathbf{r},\mathbf{v},C)$. If $\tau<t_1$ is the largest time at which $\mathbf{r}(\tau)\in \Phi$, let $(H,I_{\tau+1})$ be the flipping rotator model with the initial condition $I_{\tau+1}=(\mathbf{r}(\tau+1),\mathbf{v}(\tau+1),C(\tau+1))$. Then $\Phi$ is a reflecting structure of $(H,I_{\tau+1})$ and proposition \ref{prop:reflect} implies that $\mathbf{r}(t_2+t)=\mathbf{r}(t_1-t)$ for $0\leq t \leq t_1-\tau-1$.

Moreover, for $t=1-\tau-1$ the proof of proposition \ref{prop:reflect} implies that $\mathbf{r}(t_2+t_1-\tau-1)=\mathbf{r}(\tau+1)$ and $\mathbf{v}(t_2+t_1-\tau-1)=-\mathbf{v}(\tau)$. From equation \ref{eq:1} it then follows that
\begin{equation*}
\mathbf{r}(t_2+t_1-\tau)=\mathbf{r}(\tau+1)-\mathbf{v}(\tau)=\mathbf{r}(\tau),
\end{equation*}
which completes the proof.
\end{proof}

An example of a semi-reflector and its effect on the particle is given in the following.

\begin{example}\label{ex:2} \textbf{(Semi-Reflector Formation)}
Consider the trajectory of the particle shown in figure \ref{fig:4}(b), consisting of the solid, dashed, and dash-dotted lines, respectively. Here, $\Phi=T[7,31]$ is a semi-reflecting structure and $\tau=2$ is the largest time at which $\mathbf{r}(\tau)\in\Phi$ and $\tau<7$. Proposition \ref{prop:semi} then implies that $\mathbf{r}(31+t)=\mathbf{r}(7-t)$ for $0\leq t\leq 7-\tau$. However, as can be seen in the figure, $\mathbf{r}(38-\tau+1)\neq\mathbf{r}(\tau-1)$, so that beyond the time $t=38-\tau$, the particle no longer reverses its trajectory. Therefore, there is no guarantee that the particle will ever return to its initial position only that it will return to $\mathbf{r}(\tau)$.
\end{example}


Returning to our investigation of reflecting structures, once the particle encounters one of these structure, it will make its way back to its initial position. However, the particle's motion does not stop at that lattice site, but continues on. Therefore, the particle may continue to encounter other reflecting or semi-reflecting structures.

In the remainder of this section we study how a number of reflecting and semi-reflecting structures can interact to cause either: (i) the particle to have a periodic trajectory, (ii) annihilate each another, or (iii) form a combination of (i) and (ii).

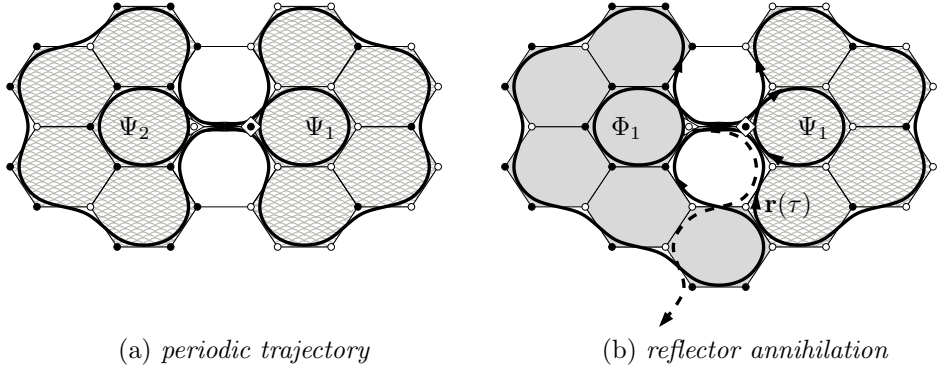
\begin{figure}
    \begin{overpic}[scale=.28]{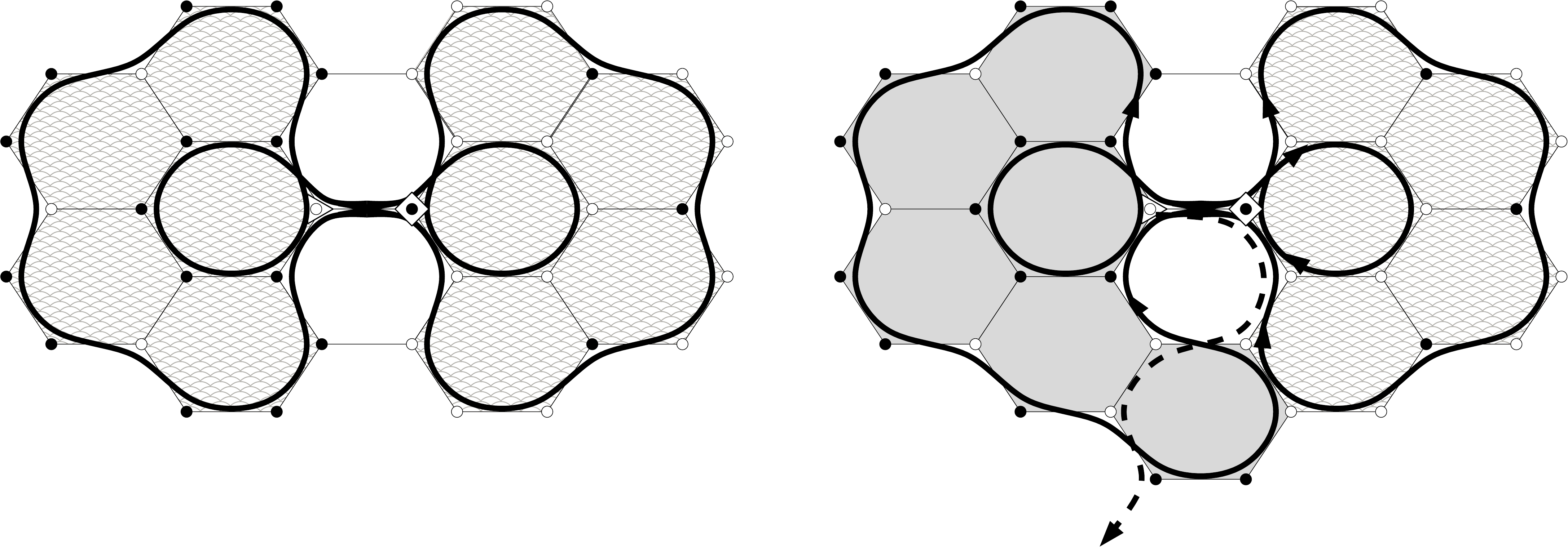}
    \put(32,21){$\Psi_1$}
    \put(12,21){$\Psi_2$}

    \put(85,21){$\Psi_1$}
    \put(65,21){$\Phi_1$}
    \put(81.5,12.75){$\tiny\mathbf{r}(\tau)$}

    \put(12,-3){(a) \emph{periodic trajectory}}
    \put(64,-3){(b) \emph{reflector annihilation}}
    \end{overpic}
    \vspace{0.1in}
\caption{A periodic trajectory of the particle, consisting of the two reflecting structures $\Psi_1$ and $\Psi_2$, is shown in (a). The particle, with initial velocity to the right, first encounters the reflecting structure $\Psi_1$, which reflects it towards $\Psi_2$. The particle, subsequently, becomes trapped between the two reflectors. Here, the particle's direction is not indicated, since the particle traverses each lattice bond of $\Psi_1$ and $\Psi_2$ in both directions. In (b), the annihilation of the reflecting structure $\Psi_1$ by the semi-reflecting structure $\Phi_1$ is shown. Here, the particle stops reversing its trajectory at the site $\mathbf{r}(\tau)$, indicated by the dashed line.}\label{fig:5}
\end{figure}

We begin by considering the case in which the particle first encounters the reflecting structure $\Psi_1=T[t_1,t_2]$, then the reflecting structure $\Psi_2=T[t_3,t_4]$ (cf. figure \ref{fig:5}). In this case, according to proposition \ref{prop:reflect}, after the particle leaves $\Psi_2$ it will reverse its trajectory and pass again through $\Psi_1$, but in the opposite direction. The reason the particle reverses it trajectory through $\Psi_1$ is that, the particle's initial trajectory through $\Psi_1$, causes the scatters of $\Psi_1$ to change orientation. In this sense, the particle \emph{transforms} $\Psi_1$ as it passes through it.

A consequence of this transformation is that, when the particle returns to this transformed version of $\Psi_1$, it will reverse its original path through this structure. This second, but reversed sequence of positions through $\Psi_1$ is, by definition, the semi-reflecting structure $\hat{\Psi}_1=T[t_5,t_6]$ for some $t_5,t_6>t_4$. Based on proposition \ref{prop:reflect}, this structure, which we call the \emph{transform} of $\Psi_1$, has the property that
\begin{equation}\label{eq:times}
\mathbf{r}(t_6+t)=\mathbf{r}(t_1-t) \ \text{for} \ 0\leq t\leq t_1.
\end{equation}

That is, a transform $\hat{\Psi}_1$ of a reflecting structure $\Psi_1$ is again a reflecting structure in that, after the particle exits the transform $\hat{\Psi}_1$, the particle will immediately reverse its trajectory back to its initial position in exactly the same way it did when it exited the original untransformed reflecting structure $\Psi_1$ (see example \ref{ex:3}). This notion, that a transformed reflecting structure is still a reflecting structure, is summarizes the following proposition. The proof of this proposition follows directly from equation \ref{eq:times}, which is itself a direct consequence of proposition \ref{prop:reversal}.

\begin{proposition}\label{prop:reversal} \textbf{(Transform Property)}
If $\hat{\Psi}=T[t_3,t_4]$ is a transform of the reflecting structure $\Psi=T[t_1,t_2]$, then $\mathbf{r}(t_4+t)=\mathbf{r}(t_1-t)$ for $0\leq t\leq t_1$.
\end{proposition}

Proposition \ref{prop:reversal} describes an important general consequence of the particle's motion, when it moves through a reflecting structure $\Psi$. As the particle passes through $\Psi$, it modifies (flips) each of the structure's scatterers so that, if the particle were to return, its motion through the structure would be reversed. In this way, the particle's motion through $\Psi$ causes $\Psi$ to change into its \emph{transform} $\hat{\Psi}$, which we denote by writing $\Psi\mapsto\hat{\Psi}$. This type of reflector transformation is illustrated in the following example.

\begin{example}\label{ex:3} \textbf{(Reflector Transformation and Periodicity)}
Consider the particle's motion in the LLG shown in figure \ref{fig:5}(a). Here, the particle first encounters the reflecting structure $\Psi_1=T[1,23]$, which causes the transformation $\Psi_1\mapsto\hat{\Psi}_1$. The particle is then reflected back through its initial position and towards the second reflecting structure $\Psi_2=T[24,46]$. Once the particle passes through $\Psi_2$, causing $\Psi_2\mapsto\hat{\Psi}_2$, it returns to its initial position through the transform $\hat{\Psi}_1$. This passage through $\hat{\Psi}_1$, therefore, causes the transformed reflector $\hat{\Psi}_1$ to transform back to the original reflecting structure $\Psi_1$, i.e. $\hat{\Psi}_1\mapsto\Psi_1$. The particle then continues on through the transform $\hat{\Psi}_2$, causing this reflector to transform back to $\Psi_2$ and the particle to arrive at its initial position at time $t_p=92$.

The particle has then visited each lattice site of $\Psi_1$, $\Psi_2$, and each point in between, an even number of times at time $t_p$, implying $C(t_p)=C$. Additionally, both $\mathbf{r}(t_p)=\mathbf{r}$ and $\mathbf{v}(t_p)=\mathbf{v}$. Since $(\mathbf{r}(t_p),\mathbf{v}(t_p),C(t_p))=(\mathbf{r}(0),\mathbf{v}(0),C(0))$, then at time $t_p$ the system is in the same state it was at time $0$. Therefore, the particle will repeat this same motion for all $t\geq 0$. That is, the particle's motion in this LLG is periodic with period $t_p=92$.
\end{example}

Thus, as demonstrated in this example, if the particle encounters two reflecting structures $\Psi_1$ and $\Psi_2$, where the particle first encounters $\Psi_1$, and does not return to $\Psi_1$ before encountering $\Psi_2$, its motion will be periodic (see theorem \ref{thm:1} and figure \ref{fig:5}(a)). This is so, because the particle will first move through $\Psi_1$ and then $\Psi_2$, causing $\Psi_1\mapsto\hat{\Psi}_1$ and $\Psi_2\mapsto\hat{\Psi}_2$. After this, the particle will reverse its motion through these two structures, moving first through the transform $\hat{\Psi}_1$ and then through $\hat{\Psi}_2$, causing $\hat{\Psi}_1\mapsto\Psi_1$ and $\hat{\Psi}_2\mapsto\Psi_2$. The result is one period of the particle's motion. The particle will then continue to repeatedly pass through both reflecting structures, in this same order. In this sense, these structures never lose their ability to \emph{limit} the particle's motion, i.e. they remain reflecting structures for all time, and are the reason for the particle's periodic motion.

This seems to suggest that, whenever a particle encounters two different reflecting structures, its trajectory will be periodic. This, however, is not always the case. Another possibility is that, before the second reflector is encountered, the first reflector is annihilated. This possibility is considered in the following example.

\begin{example}\label{ex:4} \textbf{(Reflector Annihilation)}
Consider the particle's motion in the $(H,I)$ model, as is shown in figure \ref{fig:5}(b). The particle's trajectory in this figure is similar to the particle's trajectory in figure \ref{fig:5}(a), since the particle first encounters the reflecting structure $\Psi_1=T[1,23]$, which reflects the particle back towards the second structure $\Phi_1=T[24,50]$. Here, $\Phi_1$ is not a reflecting structure but rather a semi-reflecting structure, since $\Phi_1\cap T[0,49]\neq \emptyset$. According to proposition \ref{prop:semi}, there must then be lattice site $\mathbf{r}(\tau)$, at which the particle may stop reversing its trajectory, after encountering  $\Phi_1$. In this example, this is indeed the case, as can be seen in figure \ref{fig:5}(b), where the particle's trajectory following $\Phi_1$ is indicated by a dashed line.

Based on this example, we observe that if the particle encounters first a reflector then second a semi-reflector, this is not sufficient to cause the particle's trajectory to become periodic. This, by itself, should not be surprising, since a semi-reflector may only cause the particle to reverse its trajectory up to a certain point in time (cf. example \ref{ex:2}). What is perhaps surprising, though, is that after the semi-reflector $\Phi_1$ intersects $\Psi_1$, $\Psi_1$ can no longer function as a reflector. That is, the semi-reflector $\Phi_1$ has destroyed $\Psi_1$'s ability to reverse the particle's trajectory back to its initial position.

\begin{figure}
    \begin{overpic}[scale=.245]{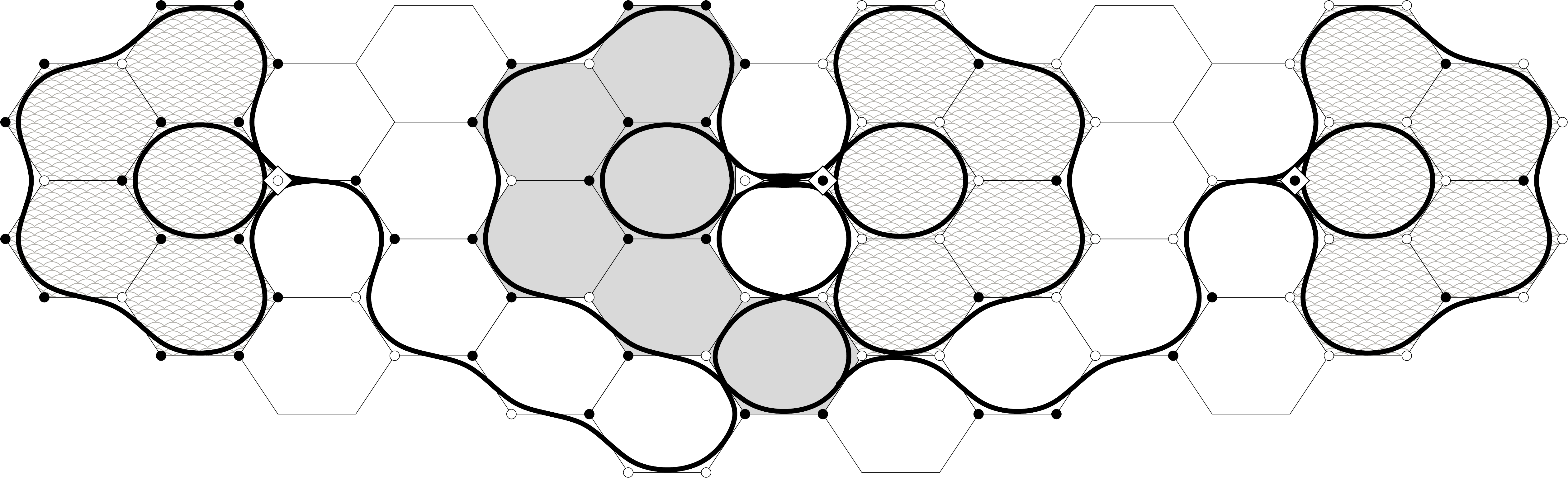}
    \put(40,18){$\Phi_1$}
    \put(10,18){$\Psi_2$}
    \put(87,18){$\Psi_3$}
    \put(57,18){$\Psi_1$}
    \put(48.75,7){$\tiny\mathbf{r}(\tau)$}
    \end{overpic}
\caption{The particle first encounters the reflecting structure $\Psi_1$, which is annihilated by the semi-reflecting structure $\Phi_1$. The particle then encounters the reflecting structures $\Psi_2$ then $\Psi_3$, which cause the particle's motion to become periodic.  The particle's direction is not indicated, since the particle traverses each lattice bond of its periodic trajectory in both directions.}\label{fig:6}
\end{figure}
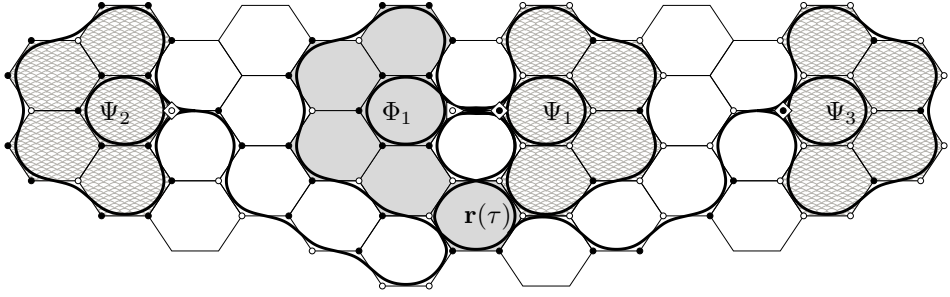

To see this, suppose that after the particle leaves $\Psi_1\cup\Phi_1$, it encounters a second reflecting structure $\Psi_2$, as is shown in figure \ref{fig:6}. If the reflecting structure $\Psi_1$ were, in fact, still acting as a reflector, after being annihilated, then the particle will become trapped by the reflectors $\Psi_1$ and $\Psi_2$, as in example \ref{ex:3}. However, the particle does not become trapped but, instead, its trajectory deviates at the lattice site $\mathbf{r}(\tau)$ from its previous trajectory, (cf. figure \ref{fig:6}). Therefore, when the particle's trajectory intersects an existing reflecting structure, the result is that the reflector's key property, of being able to reflect the particle, is lost. In this case we say that, the reflector has been \emph{annihilated}.

On the other hand, if, after the particle leaves $\Psi_1\cup\Phi_1$ for the second time, it encounters the reflecting structure $\Psi_3$, as shown in figure \ref{fig:6}, then the particle's trajectory does indeed become trapped between the reflectors $\Psi_2$ and $\Psi_3$. That is, if the particle has annihilated a reflector, the particle can still continue creating more reflecting structures, as it moves through the lattice and then eventually become trapped between two different reflectors.
\end{example}

In the following, we formally define the notion of a \emph{reflector annihilation}.

\begin{definition}
Suppose $\Psi=T[t_1,t_2]$ is a reflecting structure of $(H,I)$. Then the particle is said to \emph{annihilate} $\Psi$ at time $\tau>t_2$, if the particle:\\
(a) does not encounter a reflecting structure at any time $t\in(t_2,\tau]$, where\\
(b) $\tau$ is the first time that $\mathbf{r}(\tau)\in\Psi$.
\end{definition}

If the reflector $\Psi$ is annihilated at time $\tau>0$, then it is no longer a reflecting structure beyond this point in time. This notion allows us to prove the following theorem, which can be summarized as follows. If there are two reflecting structures at any point in time in the $(H,I)$ model, then the particle's motion in this LLG must be periodic.

\begin{theorem}\label{thm:1}
Suppose that in the $(H,I)$ model, there are two reflecting structures $\Psi_1=T[t_1,t_2]$ and $\Psi_2=T[t_3,t_4]$ at some time $\tau\geq t_4>t_3>t_2>t_1$. Then the particle's trajectory is periodic with period $t_p=2(t_4+t_3-t_2-t_1)$.
\end{theorem}

The reason this theorem holds is that, the particle will always become trapped between any two reflectors $\Psi_1$ and $\Psi_2$ it has created, but not annihilated. Specifically, after the particle encounters $\Psi_1$, it will be reflected back through its initial position and will eventually encounter $\Psi_2$. This second reflector will then reflect the particle back towards $\Psi_1$, which will in turn reflect the particle back towards $\Psi_2$. These reflections will continue for all time, in effect trapping the particle between these two structures. This interplay of reflecting structures is the main tool used to prove theorem \ref{thm:1}.

\begin{proof}
Suppose $\Psi_1=T[t_1,t_2]$ and $\Psi_2=T[t_3,t_4]$ are both reflecting structures of $(H,I)$ at some time $\tau$ where $I=(\mathbf{r},\mathbf{v},C)$ and $\tau\geq t_4>t_2$. Since $\Psi_2$ is a reflecting structure, proposition \ref{prop:reflect} implies that $\mathbf{r}(t_3-t)=\mathbf{r}(t_4+t)$ for $t\in[0,t_3]$. Thus, the particle first encounters $\Psi_1$ at time $t_1$ and then $\Psi_2$ at time $t_2$, after which it encounters the transform $\hat{\Psi}_1$ of $\Psi_1$ at time $t_4+t_3-t_2$. The particle then continues on, returning to its initial position $\mathbf{r}$ at time $t_4+t_3$.

After the particle arrives at $\mathbf{r}$, the claim is that its position satisfies the relation
\begin{equation}\label{eq:times1}
\mathbf{r}(t_4+t_3+t)=\mathbf{r}(t_2+t_1+t) \ \text{for} \ 0\leq t\leq t_3-t_2-t_1,
\end{equation}
so that the particle encounters the transform $\hat{\Psi}_2=T[t_4+2t_3-t_2-t_1,2t_4+t_3-t_2-t_1]$ of $\Psi_2$ at time $t_4+2t_3-t_2-t_1$. If this is the case, then proposition \ref{prop:reversal} implies that $\mathbf{r}(2(t_4+t_3-t_2-t_1+t))=\mathbf{r}(t_3-t)$ for $0 \leq t\leq t_3$ so that this holds, in particular, for $0\leq t\leq t_3-t_2-t_1$. Hence,
\begin{align}\label{eq:times2}
\mathbf{r}(2(t_4+t_3-t_2-t_1))&=\mathbf{r}(t_3+t_2)=\mathbf{r},\\
\label{eq:times3}\mathbf{v}(2(t_4+t_3-t_2-t_1))&=\mathbf{v}(t_3+t_2)=\mathbf{v}.
\end{align}

To verify \eqref{eq:times2}, note that proposition \ref{prop:reflect} implies that $\mathbf{r}(t_2+t_1)=\mathbf{r}(t_4+t_3)$, $\mathbf{v}(t_2+t_1)=\mathbf{v}(t_4+t_3)$, and $C_{\overline{\Psi_1\cup\Psi}_2}(t_2+t_1)=C_{\overline{\Psi_1\cup\Psi}_2}(t_4+t_3)$. This last equation follows from the fact that the particle has visited each lattice site, except those of the reflecting structures an even number of times, by the times $t_2+t_1$ and $t_4+t_3$. Given that $\mathbf{r}(t_2+t_1+t),\mathbf{r}(t_4+t_3+t)\notin \Psi_1\cup\Psi_2$ for $0\leq t\leq t_3-t_2-t_1$, then \eqref{eq:times1} holds, which in turn implies that \eqref{eq:times2}-\eqref{eq:times3} hold.

Note that each lattice site of $\overline{\Psi_1\cup\Psi}_2$ has been visited an even number of times at time $2(t_4+t_3-t_2-t_1)$. Moreover, the particle has passed through $\Psi_1$, $\hat{\Psi}_1$, $\Psi_2$, and $\hat{\Psi_2}$ exactly once by this time, so that each lattice site of $\Psi_1\cup\Psi$ has been visited an even number of times at time $2(t_4+t_3-t_2-t_1)$. Therefore, $C(2(t_4+t_3-t_2-t_1))=C$, which implies, together with equations \eqref{eq:times2}-\eqref{eq:times3}, that the particle has a periodic trajectory with period $t_p=2(t_4+t_3-t_2-t_1)$.
\end{proof}

This proof of theorem \ref{thm:1} explains why the particle in the $(H,I)$ model becomes trapped between two reflecting structures. What is fundamental to this process is, that the scattering rule in this LLG is \emph{deterministic}. That is, a right (left) rotator \emph{always} rotates the particle's velocity to its right (left) and then flips its orientation. If this were to happen, instead, with some probability $p<1$, then any reflecting structure $\Psi$ could only have the reflecting property given in proposition \ref{prop:reflect} with some probability less than 1. In this case, the probability that the particle remains trapped between two reflectors will decrease to zero, as $t\rightarrow\infty$, in which case theorem \ref{thm:1} would not hold. Thus, the deterministic nature of the scattering rule is essential for the particle's periodic dynamics in the $(H;p)$ model.

Returning to the result  of theorem \ref{thm:1}, once there are two reflecting structures in a given $(H,I)$ model, the particle's dynamics is guaranteed to be periodic. Once periodic, the particle cannot visit any new lattice sites and, therefore, cannot encounter any new reflecting structures. This limits the number of reflecting structures, that can exists at any time in a flipping rotator model, to two. This argument proves the following corollary, which follows directly from theorem \ref{thm:1}.

\begin{corollary}
At any time $t>0$, in the LLG $(H,I)$, there cannot be more then 2 reflecting structures .
\end{corollary}

Although two is the maximal number of reflectors that can exist in the $(H;p)$ model at any point in time, there is no limit to the number of different reflectors that a particle can encounter. For instance, in example \ref{ex:4}, the particle encounters three reflectors before becoming periodic. In fact, the reflector is possible that each time the particle creates a reflector, it is subsequently annihilated by the particle. In this way, the particle can encounter arbitrarily many reflecting structures as it moves through the lattice.

In the following section, we will consider whether, in the $(H;p)$ model, the particle is likely to encounter a reflecting structure and whether or not this structure is likely to be annihilated. What we find is that, if $p\in(0,1)$, the particle cannot indefinitely avoid encountering a reflecting structure $\Psi_1$. This will be the first step in showing that the particle, for these values of $p$, must have a periodic motion. The second step is to show, that it is also unlikely that the reflecting structure $\Psi_1$ is annihilated before the particle encounters a second reflecting structure $\Psi_2$.

\section{Self-Limiting Motion in the $(H;p)$ Model}\label{sec:4}
In this section, our goal is to give a proof of theorem \ref{thm:2}, which states that for $p\in(0,1)$, the particle's trajectory in the $(H;p)$ model is periodic. The proof of this result relies to a large degree on the results found in section \ref{sec:3}, specifically on the notions of reflectors, semi-reflectors, and transforms. Once a proof of this theorem has been given, we will study how the particle's period depends on the parameter $p$. This will prompt us, in section \ref{sec:5}, to investigate how the particle's dynamics abruptly changes as $p$ approaches 0 and 1.

To prove theorem \ref{thm:2}, we will first show that the flipping rotator model $(H,I)$ is \emph{invertible}, i.e. its equations of motion are time-reversible. This time-reversibility will then be used to prove that the particle in the $(H,I)$ model, for \emph{any} initial condition $I$, has a periodic trajectory if and only if it stays in a finite subset of the (infinite) lattice $H$ for all time.

To show that the particle's equations of motion  \eqref{eq:1}--\eqref{eq:3} can be reversed, we note that these equations describe the particle's motion in \emph{forward time}. That is, given $\mathbf{r}(t)$, $\mathbf{v}(t)$, and $C(t)$ we can compute each of these quantities at time $t+1$.

In the following proposition, these quantities in \emph{reverse time} are shown to exist, i.e. given $\mathbf{r}(t)$, $\mathbf{v}(t)$, $C(t)$ these quantities can be found at time $t-1$, (see equations \eqref{eq:1.1}--\eqref{eq:3.1}). The fact that these equations exist implies that, the particle's motion is time-reversible. Therefore, it is possible to recover the past states of the $(H,I)$ model based only on its present state. This summarizes the following proposition.

\begin{proposition}\label{prop:reverse}\textbf{(Time-Reversal)}
For the initial condition $I=(\mathbf{r},\mathbf{v},C)$, the particle's time-reversed equations of motion in the $(H,I)$ model are given by
\begin{align}
\mathbf{r}(t-1)&=\mathbf{r}(t)-R\big[C(t,\mathbf{r}(t))\big]\mathbf{v}(t), \label{eq:1.1}\\
\mathbf{v}(t-1)&=R\big[C(t,\mathbf{r}(t))\big]\mathbf{v}(t), \label{eq:2.1}\\
C(t-1,\mathbf{h})&=
\begin{cases}
-C(t,\mathbf{h}) &  \ \ \text{if} \ \  \mathbf{h}=\mathbf{r}(t)\\
\hspace{0.1in} C(t,\mathbf{h}) & \ \ \text{otherwise},\label{eq:3.1}
\end{cases}
\end{align}
for $t\geq1$.
\end{proposition}

The proof of this proposition is based on the following observation. In the $(H,I)$ model, if one knows the particle's velocity $\mathbf{v}(t)$ at time $t$, then there are only two possibilities for what the particle's velocity $\mathbf{v}(t-1)$ at time $t-1$ could have been. If the particle encounters a right rotator at time at time $t$, then the particle's velocity $\textbf{v}(t-1)$ will be one of these two possibilities. If the particle encounters a left rotator at time $t$ then $\mathbf{v}(t-1)$ will be the other. Since it is possible to uniquely recover $\mathbf{v}(t-1)$, based on the type of scatterer the particle encounters at time $t$, it is possible to uniquely determine the particles position $\mathbf{r}(t-1)$ at time $t-1$. Therefore, it is possible not only to know the particle's future trajectory but also its past. A proof of proposition \ref{prop:reverse} is the following.

\begin{proof}
Let $I$ be the initial condition $I=(\mathbf{r},\mathbf{v},C)$ and let $\tau\geq0$. Because of the flipping motion of the scatterers, i.e. equation \eqref{eq:3}, we have
$$R\big[C(\tau+1,\mathbf{r}(\tau+1))\big]=R\big[-C(\tau,\mathbf{r}(\tau+1))\big]= R^{-1}\big[C(\tau,\mathbf{r}(\tau+1))\big].$$
Using this together with equation \eqref{eq:2}, we have
\begin{equation}\label{eq:proof1}
\mathbf{v}(\tau)=R^{-1}\big[C(\tau,\mathbf{r}(\tau+1))\big]\mathbf{v}(\tau+1) =R\big[C(\tau+1,\mathbf{r}(\tau+1))\big]\mathbf{v}(\tau+1).
\end{equation}
Similarly, equation \eqref{eq:1} implies that
\begin{equation}\label{eq:proof2}
\mathbf{r}(\tau)=\mathbf{r}(\tau+1)-\mathbf{v}(\tau).
\end{equation}
Moreover, for each $\mathbf{h}\in\mathbb{H}$ we have
\begin{equation}\label{eq:proof3}
C(\tau,\mathbf{h})
=\begin{cases}
-C(\tau+1,\mathbf{h}) &  \ \ \text{if} \ \  \mathbf{h}=\mathbf{r}(\tau+1)\\
\hspace{0.1in} C(\tau+1,\mathbf{h}) & \ \ \text{otherwise,}
\end{cases}
\end{equation}
since only the scatterer at $\mathbf{h}=\mathbf{r}(\tau+1)$ changes orientation at time $\tau+1$.

Replacing $\tau$ by $t-1$ in equations \eqref{eq:proof1}--\eqref{eq:proof3}, yields equations \eqref{eq:1.1}--\eqref{eq:3.1}. Hence, equations \eqref{eq:proof1}--\eqref{eq:proof3} give the particle's time-reversed equations of motion for $t\geq 1$ completing the proof.
\end{proof}

Continuing towards our goal to prove theorem \ref{thm:2}, we will need to investigate what happens in the flipping rotator model if the particle's trajectory is bounded. That is, suppose the particle in $(H,I)$ remains in the subset $\Omega$ of $H$ for all time. If $\Omega$ is \emph{bounded}, i.e. finite, there must be two times $0\leq t_1<t_2$ at which $\mathbf{r}(t_1)=\mathbf{r}(t_2)$, $\mathbf{v}(t_1)=\mathbf{v}(t_2)$, and $C(t_1)=C(t_2)$. The reason for this is that, because of the discrete nature of the lattice, the particle can only assume a finite number of positions and velocities on $\Omega$. Moreover, there are only a finite number of scattering configurations possible on $\Omega$. Therefore, at some time $t_2$, the particle's position, velocity, and the configuration of scatterers on the lattice must be the same as at some previous point in time $t_1<t_2$. To summarize, if the particle's motion is bounded, its motion is either periodic or eventually periodic.

Formally, the particle's motion in $(H,I)$ is said to be \emph{eventually periodic} with \emph{period} $t_p<\infty$, if there is a $T>0$, such that $\mathbf{r}(t)=\mathbf{r}(t+t_p)$ for all $t\geq T$. Importantly, if $T=0$, we do not consider the particle's motion to be eventually periodic, since it is then \emph{periodic} with period $t_p$.

If the particle, in the LLG $(H,I)$, has either a periodic or eventually periodic motion, its trajectory must be bounded. Conversely, using the time-reversal property, i.e. proposition \ref{prop:reverse}, it follows that, if the particle has a bounded trajectory, then its motion must be periodic (see for instance \cite{Brin2002}, page 2). Hence, the particle's motion in the LLG $(H,I)$ cannot be eventually periodic.

This argument is enough to prove the following proposition, which states that bounded motion and periodic motion are identical in the $(H,I)$ model. That is, the particle remains in a finite subset of the honeycomb lattice only if it has a periodic trajectory.

\begin{proposition}\label{prop:bound}\textbf{(Boundedness Property)}
The particle's trajectory in $(H,I)$ is periodic if and only if the particle's motion is bounded.
\end{proposition}

With proposition \ref{prop:bound} in place, we are now in a position to prove theorem \ref{thm:2}, which says that the particle in the $(H;p)$ model will, with probability 1, have a periodic motion if $p\in(0,1)$. Our strategy for proving this, is to show that, if the particle does \emph{not} have a periodic, but rather an unbounded trajectory, then the particle will eventually encounter two reflecting structures, neither of which is annihilated. However, this will imply, via theorem \ref{thm:1}, that the particle's trajectory is in fact periodic, contradicting our original assumption that it had an unbounded trajectory. This contradiction will be enough to prove that we will always observe periodic motion in the $(H;p)$ model if $p\in(0,1)$. This summarizes the proof we now give of theorem \ref{thm:2}.

\begin{proof}
Let $\Omega_1\subset H$ be a single hexagon of the hexagonal lattice that contains the origin. For each $n\geq 1$, let $\Omega_{n+1}$ be a collection of hexagons, such that $\Omega_n\subset\Omega_{n+1}\subset H$, where there are exactly five hexagons between the boundary $\partial\Omega_{n}$ of $\Omega_n$ and the boundary $\partial\Omega_{n+1}$ of $\Omega_{n+1}$, as is shown in figure \ref{fig:9.1}. Let $\Psi_1$ and $\Psi_2$ be the reflecting structures shown in this figure, where $\Psi_2$ is any one of the reflectors labeled $\Psi_2$.

For the reflectors $\Psi_1=T[t_1,t_2]$ and $\Psi_2=T[t_3,t_4]$, let $\underline{\Psi}_1=T[t_1-5,t_2+5]$ and $\underline{\Psi}_1=T[t_3-6,t_4+6]$. We call $\underline{\Psi}_1$ and $\underline{\Psi}_2$ the \emph{extensions} of the reflectors $\Psi_1$ and $\Psi_2$, which are those parts of the particle's trajectory, containing $\Psi_1$ and $\Psi_2$, respectively, in the set $\Omega_{n+1}-\Omega_n$. In particular, the lattice sites $\mathbf{r}(t_1-5)=\mathbf{r}(t_2+5)$ of $\underline{\Psi}_1$ and $\mathbf{r}(t_3-6)=\mathbf{r}(t_4+6)$ of $\underline{\Psi}_2$ are on the boundary of $\partial\Omega_n$. We call these the \emph{base} of $\underline{\Psi}_1$ and $\underline{\Psi}_2$, respectively. (These are labeled $\mathbf{r}(\tau_1)$ and $\mathbf{r}(\tau_2)$ in figure \ref{fig:9.1}, respectively.) For convenience, we say that any extended reflecting structure is a \emph{version} of $\underline{\Psi}_2$, if it is the same as $\underline{\Psi}_2$ up to translation, reflection, and (or) reflection (see figure \ref{fig:9.1}, where a number of versions of $\underline{\Psi}_2$ are shown).

For $p\in(0,1)$, suppose the particle exits $\Omega_n$ for the first time at time $\tau_1>0$. The particle will then encounter the extended reflecting structure $\underline{\Psi}_1=T[t_1-5,t_2+5]$, as is shown in figure \ref{fig:9.1}, with probability $\mathbb{P}(\underline{\Psi}_1)=p^r(1-p)^\ell>0$, where $r$ and $\ell$ are the number of left and right scatterers, respectively, that are initially in $\underline{\Psi}_1$. This is shown in figure \ref{fig:9.1}, where $\mathbf{r}(\tau_1)=\mathbf{r}(t_1-5)=\mathbf{r}(t_2+5)$.

According to proposition \ref{prop:reflect}, once the particle encounters $\Psi_1$, it returns to its initial position at time $t_1+t_2$. If the particle then remains in $\Omega_n$ for all $t\geq t_1+t_2$, its motion is bounded and therefore periodic by proposition \ref{prop:bound}. If the particle does not remain in, but exits $\Omega_n$ a second time, at time $\tau_2>\tau_1$, there are two cases. Either $\mathbf{r}(\tau_1)=\mathbf{r}(\tau_2)$ or $\mathbf{r}(\tau_1)\neq\mathbf{r}(\tau_2)$.

\begin{figure}
    \begin{overpic}[height=2in, width=4.5in]{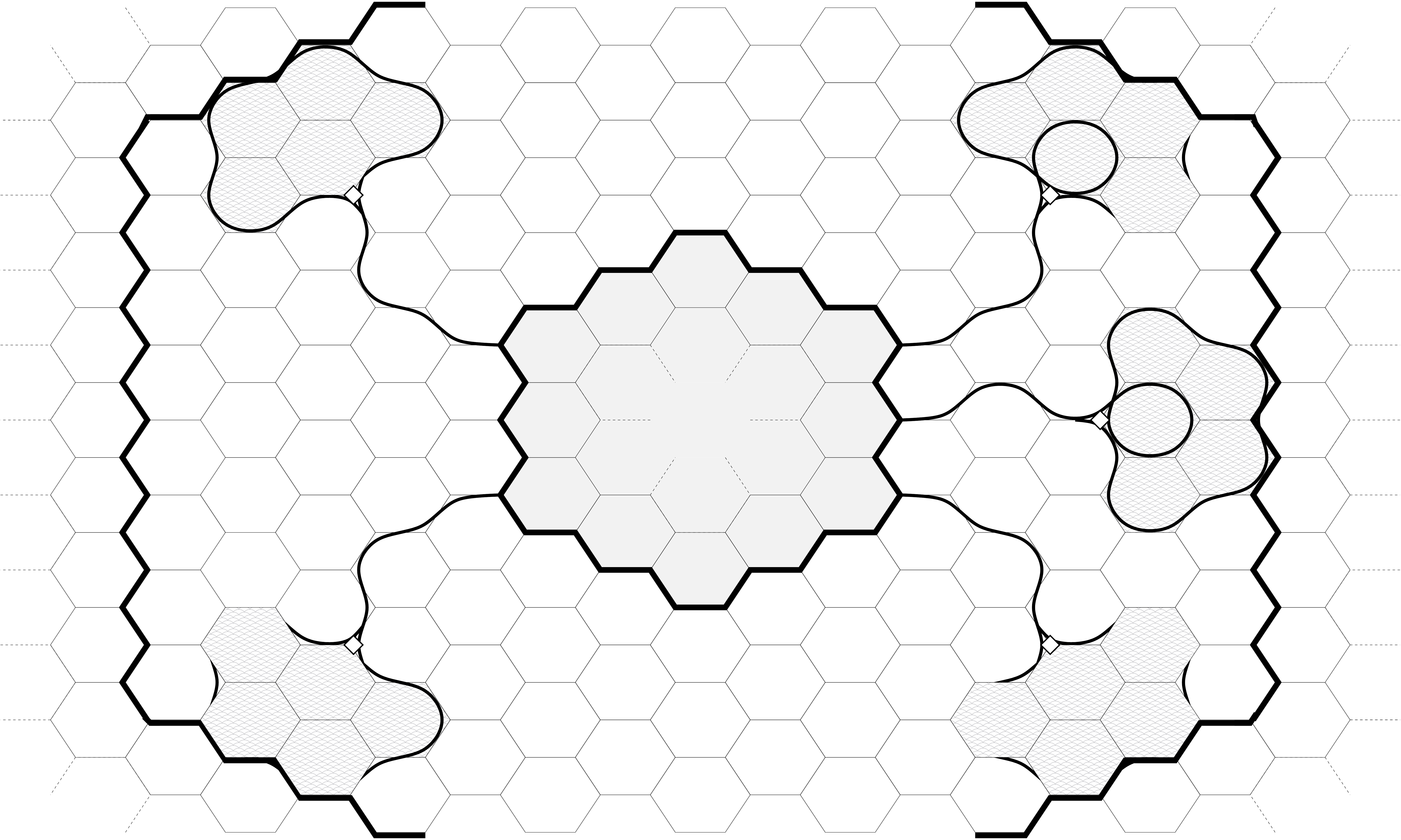}
    \put(48.5,21.5){$\Omega_n$}
    \put(46.5,-2.5){$\Omega_{n+1}$}
    \put(80.5,21.25){$\Psi_1$}
    \put(74.75,35){$\Psi_2$}
    \put(74.75,7.25){$\Psi_2$}
    \put(21.25,35){$\Psi_2$}
    \put(21.25,7.25){$\Psi_2$}

    \put(58,21.75){\tiny$\mathbf{r}(\tau_1)$}
    \put(63.25,21.35){$\bullet$}

    \put(64,27.5){\tiny$\mathbf{r}(\tau_2)$}
    \put(63.25,25.25){$\bullet$}

    \put(64,16){\tiny$\mathbf{r}(\tau_2)$}
    \put(63.25,17.25){$\bullet$}

    \put(31.5,27.5){\tiny$\mathbf{r}(\tau_2)$}
    \put(35.25,25.25){$\bullet$}

    \put(31.5,16){\tiny$\mathbf{r}(\tau_2)$}
    \put(35.25,17.25){$\bullet$}
    \end{overpic}
\vspace{0.15in}
\caption{The region $\Omega_n\subset\Omega_{n+1}$ described in the proof of theorem \ref{thm:2} is shown, along with the reflecting structures $\Psi_1=T[t_1,t_2]$ and $\Psi_2=T[t_3,t_4]$, as well as their extensions $\underline{\Psi}_1=T[t_1-5,t_2+5]$ and $\underline{\Psi}_2=T[t_3-6,t_4+6]$. A number of versions of $\underline{\Psi}_2$ are shown. These extensions are based at $\mathbf{r}(\tau_1)$ and $\mathbf{r}(\tau_2)$, respectively.}\label{fig:9.1}
\end{figure}
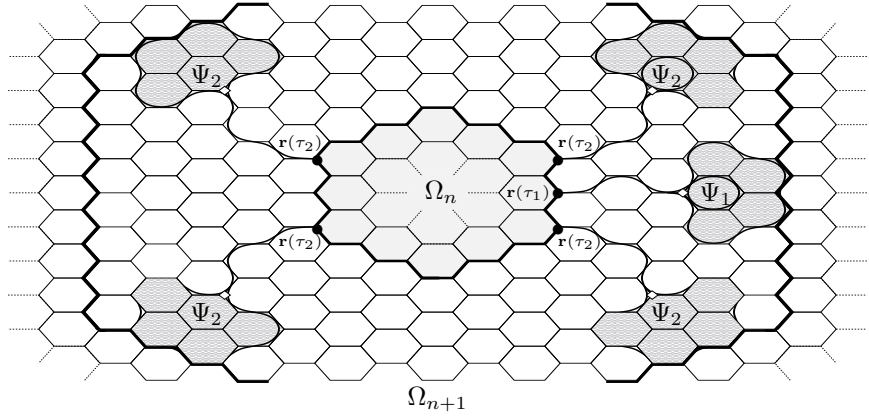

\emph{Case I}: Suppose $\mathbf{r}(\tau_1)=\mathbf{r}(\tau_2)$. Then $\mathbf{v}(\tau_1)=\mathbf{v}(\tau_2)$ and moreover the configuration $C_{\overline{\Psi_1\cup\Omega}_n}(\tau_1)=C_{\overline{\Psi_1\cup\Omega}_n}(\tau_2)$. Hence, $\mathbf{r}(\tau_1+t)=\mathbf{r}(\tau_2+t)$ for $t\in[0,t_1-\tau_1]$, where $\mathbf{r}(\tau_2+t_1-\tau_1)=\mathbf{r}(t_1)$ is the base of the nonextended reflecting structure $\Psi_1$. Therefore, at time $\tau_2+t_1-\tau_1$, the particle encounters the transformed reflector $\hat{\Psi}_1=T[\tau_2+t_1-\tau_1,\tau_2+t_2-\tau_1]$ of $\Psi_1$, which according to proposition \ref{prop:reversal} will cause the particle to reverse its trajectory back to its initial position.

In particular, at time $t_p=2(t_4+t_3-t_2-t_1)$, where $t_4=\tau_2+t_2-\tau_1$ and $t_3=\tau_2+t_1-\tau_1$, we have that $\mathbf{r}(t_p)=\mathbf{r}$, $\mathbf{v}(t_p)=\mathbf{v}$, and $C(t_p)=C$, implying that the particle's trajectory is periodic with period $t_p$. This follows, using the same argument used to prove theorem \ref{thm:1}.

\emph{Case II}: Suppose $\mathbf{r}(\tau_1)\neq\mathbf{r}(\tau_2)$. By construction, some version of the extended reflector $\underline{\Psi}_2=T[t_3-6,t_4+6]$, based at $\mathbf{r}(\tau_2)$, does not intersect any lattice site of $\underline{\Psi}_1$ (cf. figure \ref{fig:9.1}). The particle encounters this version of $\underline{\Psi}_2$ with probability $\mathbb{P}(\underline{\Psi}_2)=p^{r}(1-p)^{\ell}$, where $r$ and $\ell$ are, in this case, the number of left and right scatterers that are initially in $\underline{\Psi}_2$, respectively.

Since the extended reflecting structures $\underline{\Psi}_1$ and $\underline{\Psi}_2$ do not overlap spatially, the reflector $\Psi_1$ is not annihilated before the particle forms the second reflector $\Psi_2$. Hence, the particle must have a periodic trajectory, via theorem \ref{thm:1}. However, this happens only with some positive probability $\mathbb{P}(\Psi_2)>0$. If this does not happen, then there are two possibilities. Either the particle remains in $\Omega_{n+1}$ for all time, so that the particle's motion is periodic, or the particle exits $\Omega_{n+1}$ at some future point in time. If the latter is true, we can repeat our previous argument to show that, with some positive probability, the particle's motion will be periodic. This can be done as follows.

Let $E_n$ be the event that if the particle exits $\Omega_n$ twice, at two different lattice sites, it first encounters the extended reflecting structure $\underline{\Psi}_1\subset\Omega_{n+1}$, and later some version of $\underline{\Psi}_2\subset\Omega_{n+1}$, where $\underline{\Psi}_1\cap\underline{\Psi}_2=\emptyset$. Then
\begin{equation}\label{eq:prob}
\mathbb{P}(E_n)\geq\mathbb{P}(\underline{\Psi}_1)\times\mathbb{P}(\underline{\Psi}_2)>0,
\end{equation}
since (1) the assumption that $p\in(0,1)$ implies $\mathbb{P}(\underline{\Psi}_i)>0$ for $i=1,2$ and (2) the assumption that the scatterer's orientations are independent implies that the particle encountering $\underline{\Psi}_1$ and $\underline{\Psi}_2$ are independent events. This assumption of independence in the $(H;p)$ model, furthermore, implies that $\{E_n\}$ is a collection of independent events. Moreover, \eqref{eq:prob} implies that $\sum_{i=1}^\infty=\mathbb{P}(E_n)=\infty$. The second Borel-Cantelli lemma (see \cite{Feller86}, p. 201, for instance), therefore implies that $\mathbb{P}(\limsup_{n\rightarrow\infty}E_n)=1$ so that, with probability 1, the particle encounters two reflecting structures, neither of which is annihilated.

Therefore, either the particle becomes trapped in some $\Omega_n$ for all time and has a periodic trajectory by proposition \ref{prop:bound} or, with probability 1, at some point in time, there must be two reflecting structures in the $(H;p)$ model. In the latter case, theorem $\ref{thm:2}$ implies that the particle's trajectory is periodic. Since these two cases are the only possibilities, this completes the proof.
\end{proof}

\begin{figure}
\begin{center}
\begin{tabular}{cc}
    \begin{overpic}[scale=.19]{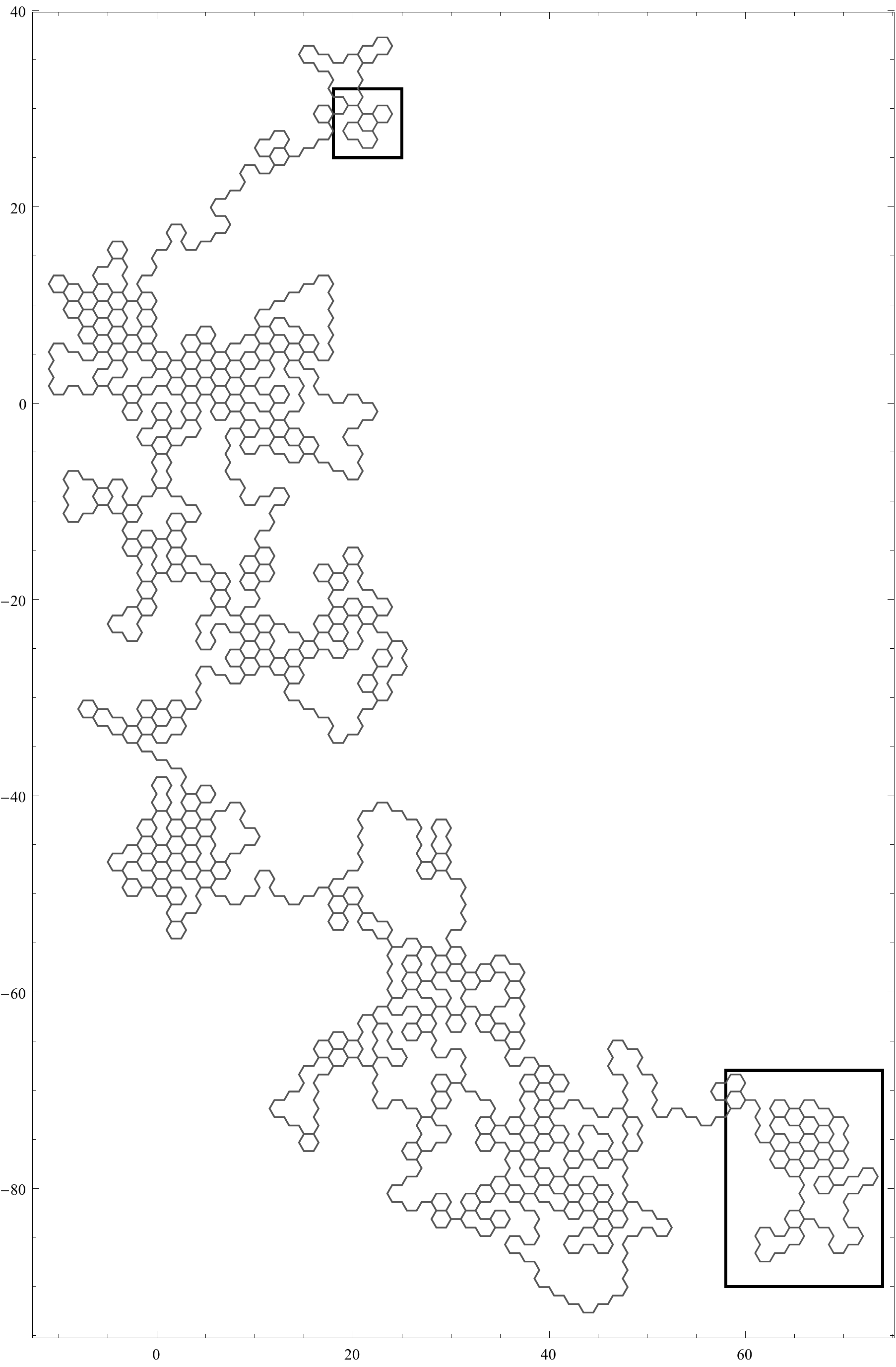}
    \put(31,89){$\Psi_1$}
    \put(55,24){$\Psi_2$}
    \put(33,-10.75){(a)}
    \end{overpic} &
    \begin{overpic}[scale=.27]{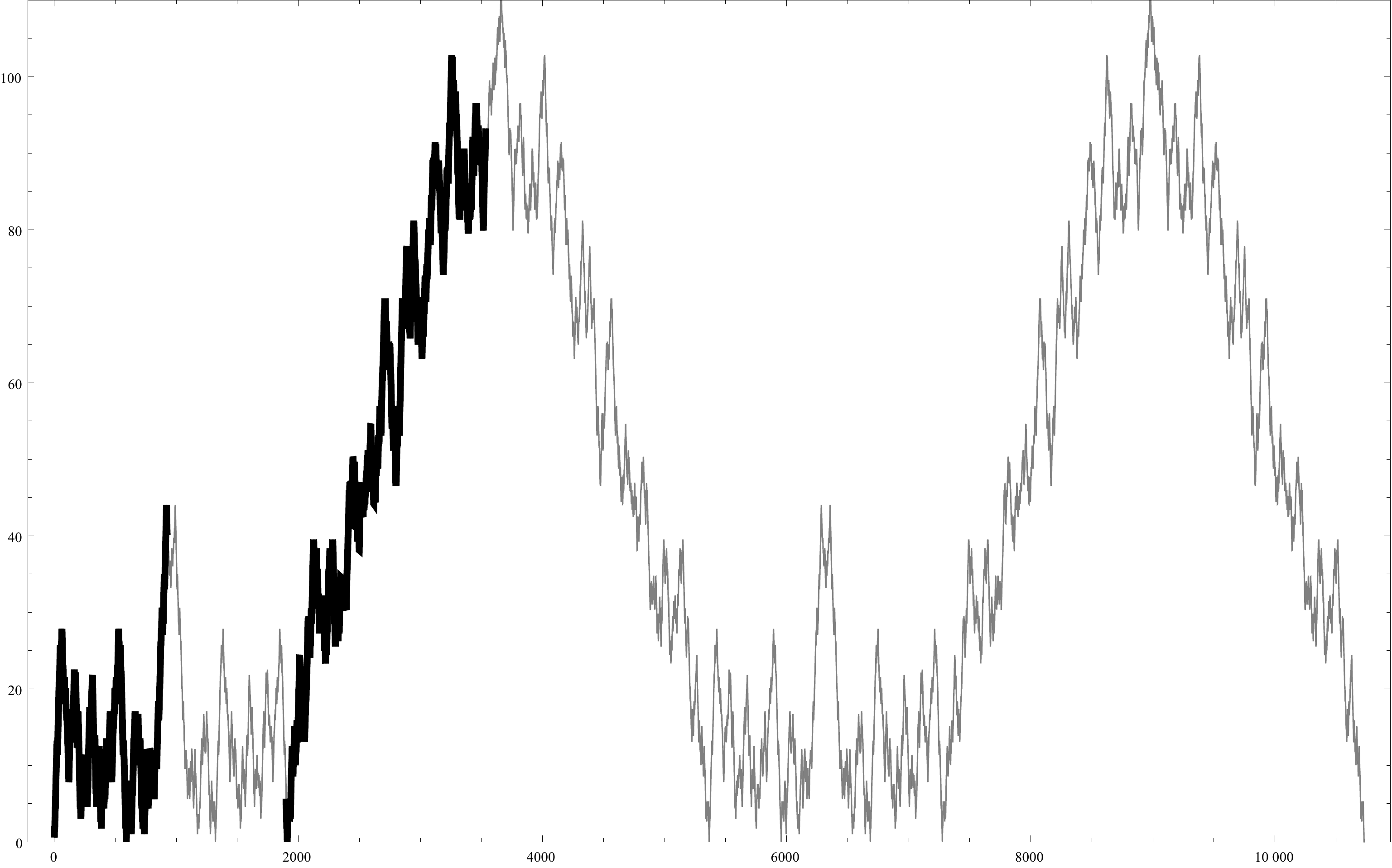}
    \put(50,-7.5){(b)}

    \put(11,0.5){$|$}
    \put(7,-3){\tiny{$t=944$}}
    \put(9.5,22){\fbox{\parbox[b][0.75em][t]{0.0075\textwidth}{\hspace{0.01in} } }}
    \put(6,28){\tiny{$\Psi_1\mapsto\hat{\Psi}_1$}}

    \put(35,0.5){$|$}
    \put(31,-3){\tiny{$t=3602$}}
    \put(80,56.45){\fbox{\parbox[b][1em][t]{0.01\textwidth}{\hspace{0.01in} } }}
    \put(40,57.5){\tiny{$\Psi_2\mapsto\hat{\Psi}_2$}}

    \put(58,0.5){$|$}
    \put(54,-3){\tiny{$t=6308$}}
    \put(56.75,22){\fbox{\parbox[b][0.75em][t]{0.0075\textwidth}{\hspace{0.01in} } }}
    \put(53,28){\tiny{$\hat{\Psi}_1\mapsto\Psi_1$}}

    \put(82,0.5){$|$}
    \put(78,-3){\tiny{$t=8966$}}
    \put(33,56.45){\fbox{\parbox[b][1em][t]{0.01\textwidth}{\hspace{0.01in} } }}
    \put(87,57.5){\tiny{$\hat{\Psi}_2\mapsto\Psi_2$}}

    \end{overpic}
\end{tabular}
\end{center}
\vspace{0.1in}
\caption{A realization of the particle's trajectory in the $(H,1/2)$ model is shown in (a), along with the reflectors $\Psi_1$ and $\Psi_2$. The particle's sequence of displacements $\{|\mathbf{r}(t)|\}_{t\leq t_p}$ for $t_p=10728$ is shown in (b), for one period of its motion. The times on the $t$-axis at which the particle encounters and transforms the reflectors $\Psi_1=T[944,998]$, $\Psi_2=T[3602,3704]$, $\hat{\Psi}_1=T[6308,6362]$, and $\hat{\Psi}_2=T[8966,9068]$ are indicated. The thick black lines indicate the times at which the particle's motion is a pseudo-random walk. Thin black lines indicate the particle's subsequent reflections back to its initial position.}\label{fig:7.7}
\end{figure}
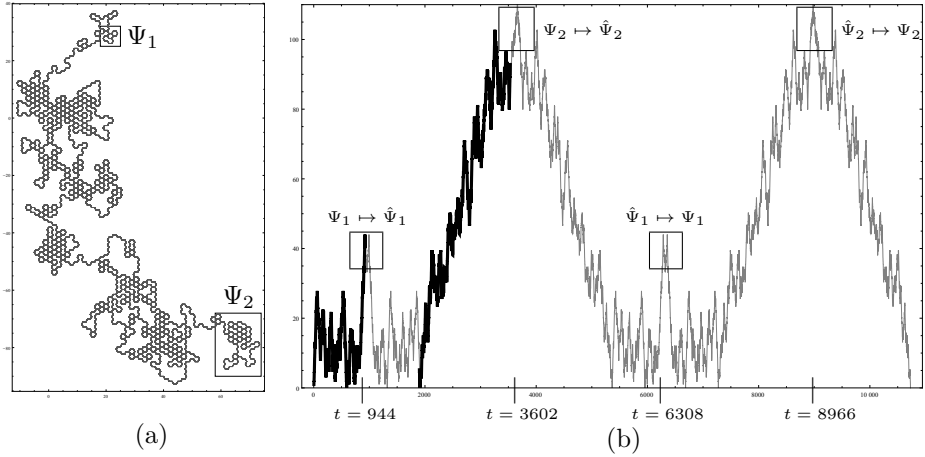

Theorem \ref{thm:2} guarantees that the particle's motion in the $(H;p)$ model will always be periodic, if $p\in(0,1)$. To illustrate how this periodicity arises, we consider a realization of the particle's trajectory in the $(H;p)$ model, shown in figure \ref{fig:7.7}, where $p=1/2$. Here, the two reflecting structures $\Psi_1$ and $\Psi_2$, that cause the particle to have a periodic trajectory, are indicated. In figure \ref{fig:7.7}(b) the particle's sequence of displacements $\{|\mathbf{r}(t)|\}_{t\leq t_p}$ are shown, where $t_p=10728$. Additionally, the times at which the particle encounters and transforms $\Psi_1$ and $\Psi_2$ are shown, as well as the times at which $\hat{\Psi}_1$ and $\hat{\Psi}_2$ are transformed back into $\Psi_1$ and $\Psi_2$, respectively.

Recall that, once the particle encounters a reflector, its future motion is then fixed, up to the point in time when it returns to its initial position (see proposition \ref{prop:reflect}). However, before it encounters a reflector, its motion can be considered to be a \emph{pseudo-random walk}, in the sense that its motion is partly random, due to the random initial configuration of the scatterers, and partly deterministic, due to the deterministic rules of motion. Hence, in figure \ref{fig:7.7}, the particle's motion is a pseudo-random walk until time t=944, when it encounters $\Psi_1$. After returning to its initial position at time $t=1942$, the particle's motion is again a pseudo-random walk until it reaches $\Psi_2$ at time $t=3602$.

Once the particles encounters this second reflector $\Psi_2$, its motion, for all future times, is completely determined by its previous motion through the lattice (see theorem \ref{thm:1}). Hence, each reflector causes the particle's motion to abruptly transition from a pseudo-random walk to a deterministic one. In fact, after the particle has encountered two reflectors, which have not been annihilated, as is the case in figure \ref{fig:7.7}(a), the particle's motion has become completely deterministic, so that the randomness of the initial configuration no longer affects the particle's motion.

Now, having established that the particle has a periodic motion in the $(H;p)$ model so long as $p\in(0,1)$, we will investigate, how the particle's period $t_p$ depends on the parameter $p$. Our motivation for doing so is to understand, how the particle's dynamics changes in the $(H;p)$ model as $p\rightarrow0,1$.

\begin{figure}
\begin{center}
\begin{tabular}{cc}
    \begin{overpic}[scale=.75]{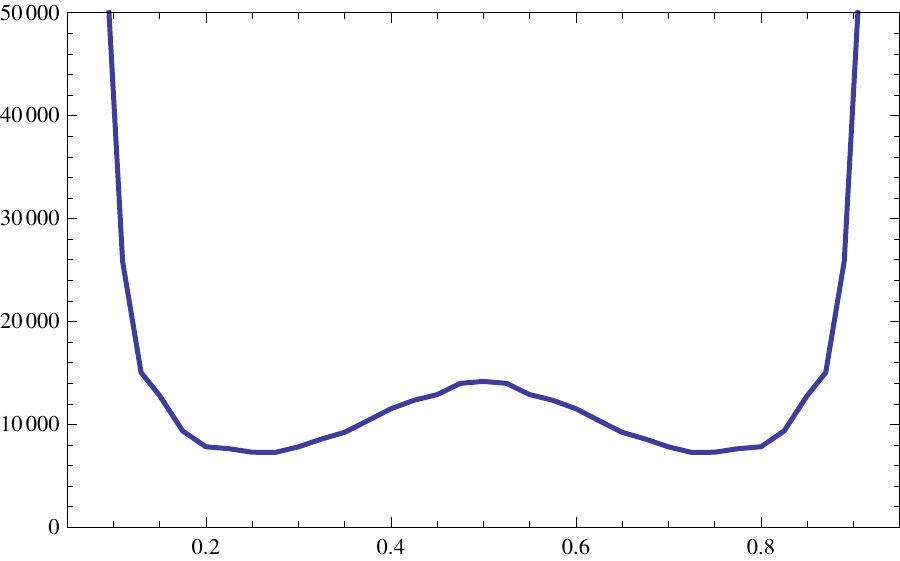}
    \put(46.5,-5){$p$-axis}
    \put(-17,31){$t_{av}(p)$}
    \put(99,0.75){\tiny{1}}
    \end{overpic}
\end{tabular}
\end{center}
\caption{The average period $t_{av}(p)$ of the particle in the $(H;p)$ model is shown for $p\in(0,1)$.}\label{fig:7.0}
\end{figure}

Using numerical simulations, the particle's average period $t_{av}(p)$, averaged over realizations of the models initial configurations, is plotted for $p\in(0,1)$ in figure \ref{fig:7.0}. What is perhaps the most striking feature of this graph is the fact, that it appears that
$$\lim_{p\rightarrow 0,1}t_{av}(p)=\infty.$$
This suggests, that at $p=0,1$ the particle's motion is non-periodic and is therefore unbounded, by proposition \ref{prop:bound}.

This raises two questions. First, what is the nature of this unbounded motion and second, how does the particle's motion make the transition from the periodic motion we observe for $p\in(0,1)$, to the unbounded motion we observe in figure \ref{fig:7.0} for $p=0,1$. Both of these questions are considered in the following section.

\section{A Discontinuous Dynamic Transition to Self-Avoiding Behavior}\label{sec:5}
In this section we investigate how the motion of the particle in the $(H;p)$ model changes as $p$ approaches 0 and 1. That is, we will investigate, how the particle's dynamics changes as the scattering configuration, i.e. the medium through which the particle moves, becomes more homogenous, until at $p=0,1$ the initial configuration consists of all left or all right rotators, respectively. What we will find is, that the particle makes a dynamic transition from the self-limiting periodic behavior, observed in section \ref{sec:4}, to both a non-periodic as well as a self-avoiding mode of motion. Specifically, for $p=0,1$, the particle's trajectory in the $(H;p)$ model is a self-avoiding walk between returns to the particle's initial position to which it will return an infinite number of times (see theorem \ref{thm:1.1}). Moreover, based on numerical simulations, the particle's trajectory is no longer periodic but unbounded (cf. figure \ref{fig:7.0}).

To describe the transition between these two types of motion, we begin by noting that for $p\in(0,1)$, the model's initial configuration is randomly generated, whereas for $p=0,1$ the model's initial configuration is deterministically generated. Specifically, for $p=0,1$ the $(H;p)$ model's initial configuration consists of either all left scatterers or all right scatterers, respectively. We note that, since both of these configurations are homogeneous, they are ordered in contrast to the random configurations considered in the previous section, for $p\in(0,1)$. As a consequence, the particle in the $(H;p)$ model for $p=0,1$ has a very different type of motion when compared with the values $p=(0,1)$.

In order to describe the dynamics in these models, we will use the notion of a \emph{self-avoiding cycle} (cf. \cite{Webb14}). Suppose that in the LLG $(H,I)$, there are two times $t_1<t_2$, such that each position $\mathbf{r}(t)$ is distinct for $t_1 \leq t< t_2$ and $\mathbf{r}(t_1)=\mathbf{r}(t_2)$. If this is the case, we say that the particle moves on the \emph{self-avoiding cycle}\footnote{We note that, sometimes these self-avoiding cycles are referred to as \emph{self-avoiding polygons} \cite{Guttmann12}.} denoted by $\gamma=T[t_1, t_2]$ from time $t_1$ to $t_2$. Moreover, we call the position $\mathbf{r}(t_1)=\mathbf{r}(t_2)$ the \emph{base} of the cycle.

To describe the difference between the particle's motion in the $(H;p)$ model for $p\in(0,1)$ and $p=0,1$, we need the following theorem, found in \cite{Webb14} (theorem 3.2). This theorem states that the particle's trajectory in the $(H,p)$ model for $p=0,1$, can be described in terms of cycles. Specifically, the particle's entire trajectory can be decomposed into a sequence of self-avoiding cycles, each of which is based at the origin.

\begin{theorem}\label{thm:1.1} \textbf{(Cyclic Decomposition Theorem \cite{Webb14})}
For $p=0,1$ there is an infinite sequence of times $\{\tau_i\}_{i\geq 0}$ in the $(H;p)$ model, with $\tau_0=0$, such that\\
(a) $\gamma_i=T[\tau_{i-1},\tau_{i}\big]$ is a self-avoiding cycle based at the origin for $i>0$; and\\
(b) each $\gamma_i$ is symmetric with respect to the line $x=1/2$.
\end{theorem}

Therefore, between returns to the origin, the particle in the $(H;0)$ and $(H;1)$ models moves through the lattice \emph{without crossing} its trajectory. In stark contrast to this, what we observe in the $(H;p)$ model, for all $p\in(0,1)$, is that the particle moves through the lattice until it encounters a reflecting structure. Then particle reverses its entire trajectory returning to each of the lattice sites it has previously visited, until it again arrives at its initial position.

A natural question then is, why is there such a difference in the particle's motion in the $(H;p)$ model for $p\in(0,1)$ and $p\in0,1$. One immediate answer is that, for each $p\in(0,1)$, there is always a positive probability that the particle will encounter a reflecting structure at some point in time $t>0$, which follows directly from the proof of theorem \ref{thm:2}. However, if $p=0,1$ then the probability for this is zero. Hence, it is not possible, in either the $(H;0)$ or $(H;1)$ model, for the particle to be ``reflected" back to its initial position by a reflecting structure. This observation is stated as the following corollary of theorem \ref{thm:2} and theorem \ref{thm:1.1}.

\begin{corollary}\label{cor:2}
No reflecting structures are possible in the $(H;p)$ model for $p=0,1$.
\end{corollary}

The physical reason that there are no reflecting structures for $p=0,1$ in this model is that, for these $p$-values, the particle's trajectory is self-avoiding away from its initial position. Since the particle must cross its own trajectory, away from its initial position, in order to create a reflector, there can then be no such structures. This is the main idea used in the following proof we give of corollary \ref{cor:2}.

\begin{proof}
For $p=0,1$, suppose that the particle in the $(H;p)$ model encounters a reflecting structure $\Psi=T[t_1,t_2]$. Part (a) of theorem \ref{thm:1.1} then implies that, the first loop $L_1=T[t_1,t_*]$ of $\Psi$ (see definition \ref{def:refstruct}) must contain the origin, since $\mathbf{r}(t_1)$ and $\mathbf{r}(t_*)$ are the same lattice site. The fact that $\mathbf{r}(0)$ is the origin and $T[0,t_1-1]\cap\Psi=\emptyset$, then leads to a contradiction of definition \ref{def:refstruct} part (c). Hence, the particle can never encounter a reflecting structure in the $(H;p)$ model, if $p=0,1$.
\end{proof}

Based on corollary \ref{cor:2}, no reflecting structures are possible in $(H;p)$ for $p=0,1$. For $p\in(0,1)$, what we observe numerically is that, the particle's trajectory in the $(H;p)$ model only becomes periodic after encountering a number of reflecting structures. This leads us to conjecture that the probability of ever encountering a reflecting structure is in fact equal to $1$ for $p\in(0,1)$. This is in contrast to what we observe in these numerical simulations for $p=0,1$, where the particle's trajectory appears to be unbounded and therefore non-periodic (cf. figure 7 in \cite{Webb14}).

To physically understand the transition between the periodic self-limiting motion and the self-avoiding motion observed in the $(H;p)$ model, we consider the case in which $p\in(0,1)$ is slightly less than 1. Then the concentration of scatterers that are initially oriented to the left is $1-p$, which is then slightly greater than 0. Since the particle in the $(H;p)$ model and the $(H;1)$ model have the same trajectory only until the particle in the $(H;p)$ model encounters a left scatterer then, over short time intervals, these two models have the same dynamics, with a high probability.

Consequently, the particle's transient, or short term dynamics in the $(H;p)$ model, for $p$ close to 1, consists, with a high probability, of a number of self-avoiding cycles, since the particle's dynamics in the $(H,;1)$ model consist entirely of self-avoiding cycles. As $p$ approaches 1, the number of self-avoiding cycles increases as the particle's trajectory in the $(H;p)$ and the $(H;1)$ model become more similar until, in the limit, the particle's entire trajectory can be decomposed into a sequence of these cycles. It is in this manner then that the particle makes a transition from the periodic self-limiting motion, described in theorem \ref{thm:2}, to the self-avoiding motion described in theorem \ref{thm:1.1}. Moreover, by symmetry, the same holds as $p$ approaches 0.

As a final remark, we note that, although the self-avoiding motion observed in the $(H;p)$ model for $p=0,1$, is very different from the periodic motion observed for $p\in(0,1)$, there is one striking similarity between the two.

For any $p\in[0,1]$ the particle in the $(H;p)$ model will always return to its initial position an infinite number of times. For $p\in(0,1)$, the reason is, that the particle has a periodic motion, so every point of its trajectory is visited an infinite number of times. For $p=0,1$, the reason for this behavior is more subtle and is described in detail in \cite{Webb14} section 6. This \emph{recurrence property} is formally stated in the following corollary, which follows immediately from theorem \ref{thm:2} and theorem \ref{thm:1.1}.

\begin{corollary}\textbf{ (Recurrence Property)}
For $p\in[0,1]$ the particle in the $(H;p)$ model returns to its initial position an infinite number of times with probability 1.
\end{corollary}

Before continuing to the next section, we recall that in the $(H;p)$ model the initial orientations of the scatterers are, as a set, a collection of i.i.d. random variables (see definition \ref{def:frmprob}). In the following sections we investigate the particle's dynamics if these orientations are \emph{not} identically distributed (see section \ref{sec:6}), and alternately, if there are correlations between these orientations, i.e. these orientations are not independent (see section \ref{sec:7}).

\section{Non-Identically Distributed Configurations}\label{sec:6}
In the $(H;p)$ model, considered in sections \ref{sec:2}--\ref{sec:5}, the model's initial configuration of scatterers was assumed to be both independent and identically distributed. In this section we consider a generalization of this model, in which the model's initial orientations $C$ are not identically distributed. What we find is that, as long as the initial orientations remain independent from each other then, under some mild conditions, the particle in this modified version of the $(H;p)$ model, will still have a periodic trajectory.

To make this precise, let $F=\{f_1,\dots,f_n\}$ where $f_i:[0,1]\rightarrow[0,1]$ for each $1\leq i\leq n$.
For this family of functions $F$, we let $(H;F(p))$ denote the flipping rotator model with initial position and  velocity $\mathbf{r}=(0,0)$ and $\mathbf{v}=(1,0)$, respectively. This model's initial configuration $C$ is the random initial configuration where,
for each $\mathbf{h}\in\mathbb{H}$, we assume that
\begin{equation}\label{assump3}
\mathbb{P}[C(\mathbf{h})=1]=f_i(p) \ \text{for some} \ f_i\in F.
\end{equation}
Furthermore, we suppose that the initial orientations of the scatterer are independent of each other, so that $C$ is a collection of independent random variables.

Observe that if $n=1$, $F=\{f_1\}$ is a single function. In that case, the collection of initial orientations $C$ in the $(H;F(p))$ model form a collection of i.i.d. random variables. In fact, if $f_1(p)=p$, then $(H;F(p))$ is the $(H;p)$ model considered in sections \ref{sec:2}--\ref{sec:5}. In contrast, if $n>1$, the collection of initial orientations in the $(H;F(p))$ model need not be identically distributed, since there is then more than one function in the set $F$.

However, for any $n\geq 1$, the particle's motion in the $(H;F(p))$ model will be periodic as long as the model's initial configuration is still \emph{randomly generated}, i.e. if each function $f_i(p)\neq 0,1$, so that none of the scatterers are initially oriented to the right or left, with probability 1.

This result can be summarized as follows. If, in a flipping rotator model, the initial orientations of the scatterers are generated both randomly and independently, then it does not matter whether each orientation is probabilistically generated in the same way, the result will be qualitatively the same: the particle will have a periodic trajectory. This is formally stated in the following corollary.

\begin{corollary}\label{cor:4}
For $p\in[0,1]$, the particle in the $(H;F(p))$ model has a periodic trajectory, with probability 1, if $f_i(p)\neq 0,1$, for each $f_i\in F$.
\end{corollary}

The proof of this result is based on the fact that, even if the initial orientations of scatterers in a given flipping rotator model are not identically distributed, reflecting structures still exist with a positive probability throughout the lattice. Eventually, the particle can and will become trapped between two of these structures.

\begin{proof}
Let $\Psi$, $\Psi_1$ and $\Psi_2$ be the reflecting structures described in the proof of theorem \ref{thm:2}. Under the assumption that $f_i(p)\neq 0,1$ for each $f_i\in F$, it follows that $\mathbb{P}(\Psi)$, $\mathbb{P}(\Psi_1)$, $\mathbb{P}(\Psi_2)>0$. From the proof of theorem \ref{thm:2}, it follows that the particle either becomes trapped in the region $\Omega_m$ for some $m\geq 1$ or encounters, with probability 1, two reflecting structures, neither of which is annihilated. Proposition \ref{prop:bound} and theorem \ref{thm:1} together then imply, that the particle's trajectory in the $(H;F(p))$ model is periodic with probability 1.
\end{proof}

We note that the only condition needed for corollary \ref{cor:4} to hold, is that $F$ is a \emph{finite} collection of functions. Then, as long as each $f_i(p)\neq 0,1$, the particle in the $(H;F(p))$ model will, with probability 1, have a periodic trajectory. That is, the result given in theorem \ref{thm:2} can be extended to the case in which, the model's initial configuration of scatterers is not identically distributed, but only independent of each other.

Additionally, we note that, according to corollary \ref{cor:4}, the scatterer's initial orientation can be distributed in nearly any way since there are no restrictions on the functions in $F$. As an example of this result, we consider the following.

\begin{figure}
\begin{center}
\begin{tabular}{ccc}
    \begin{overpic}[scale=.126]{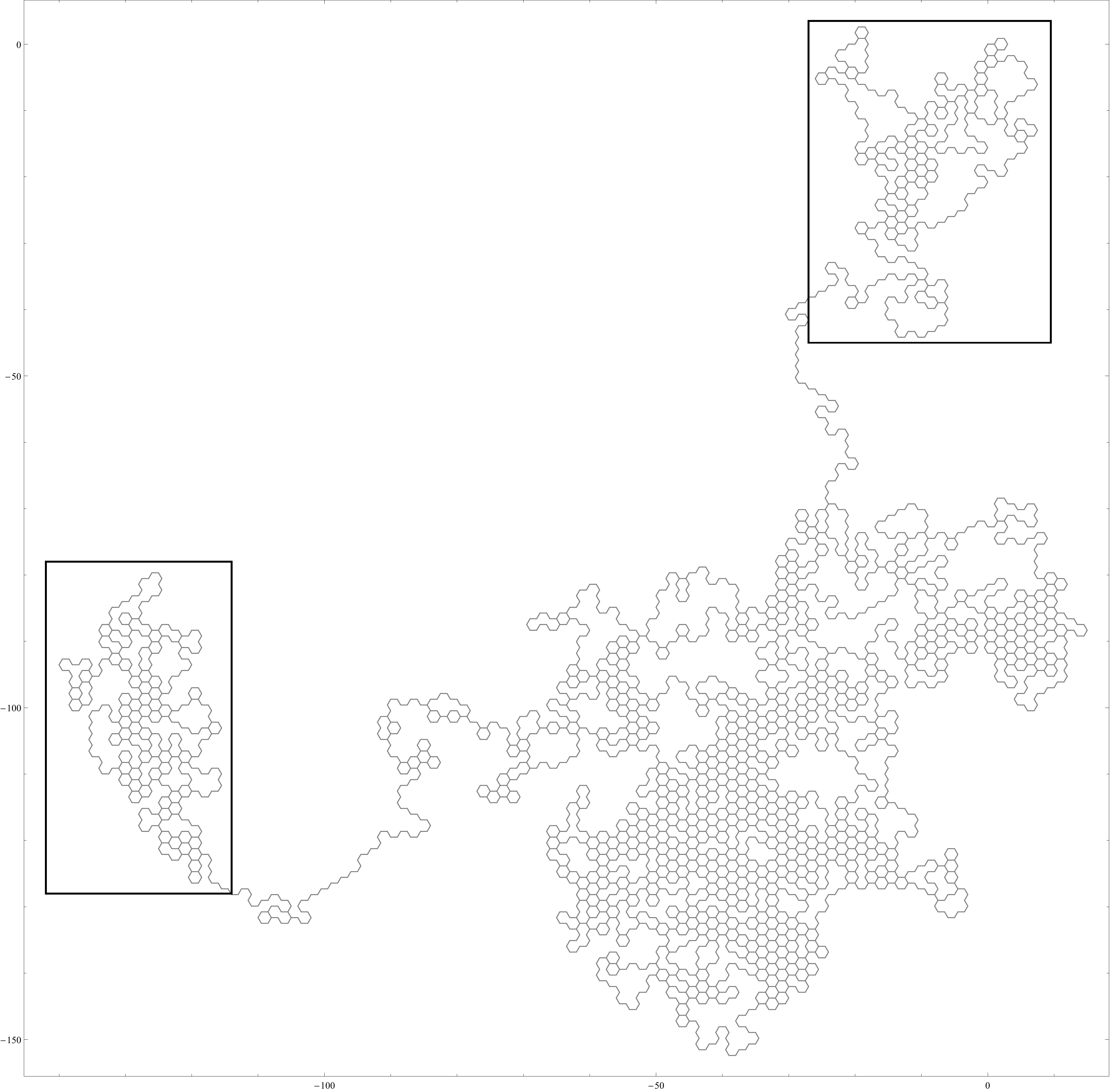}
    \put(10,11){$\Psi_2$}
    \put(82,60){$\Psi_1$}
    \put(40,-6){$p=0.2$}
    \end{overpic} &
    \begin{overpic}[scale=.15]{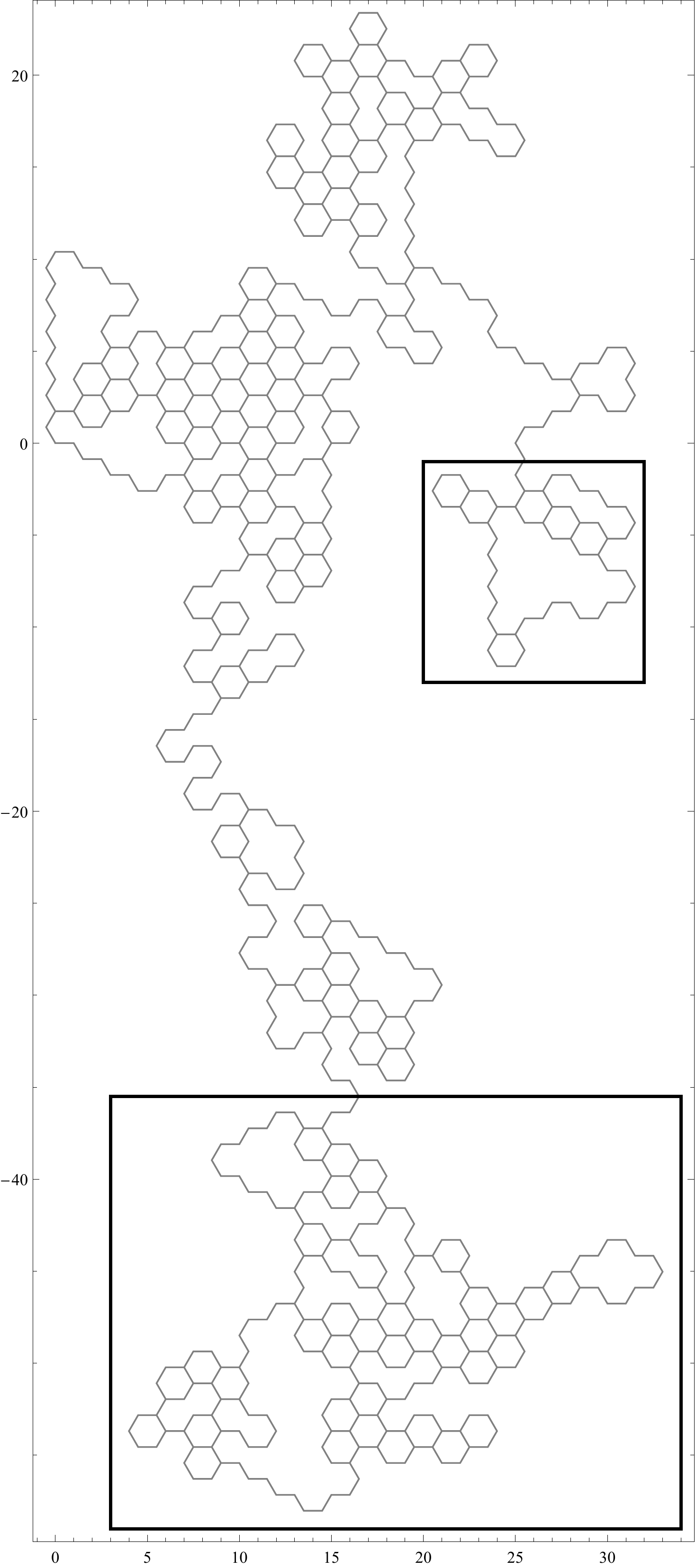}
    \put(30,32){$\Psi_2$}
    \put(32,50){$\Psi_1$}
    \put(16,-6){$p=0.5$}
    \end{overpic} &
    \begin{overpic}[scale=.24]{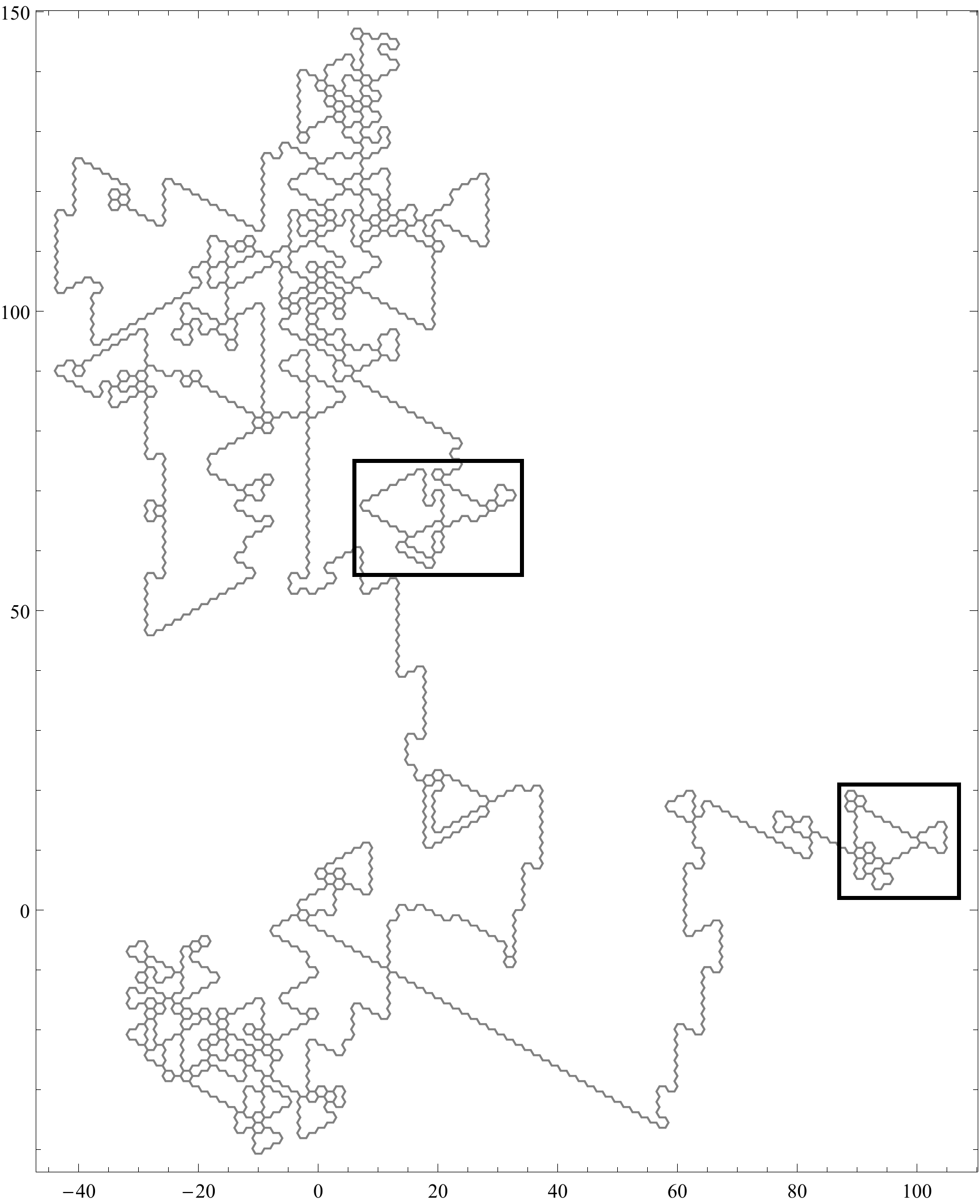}
    \put(35,46){$\Psi_2$}
    \put(71,37){$\Psi_1$}
    \put(28,-6){$p=0.7$}
    \end{overpic}
\end{tabular}
\end{center}
\caption{A realization of the particle's trajectory in the $(H;F(p))$ model, described in example \ref{ex:5}, for $p=0.2,0.5$, and $0.7$. In each case, the particle's trajectory is periodic, which for each value of $p$, is due to the formation of two reflecting structures $\Psi_1$ and $\Psi_2$. The periods are $t_p=41127,3748$, and $12450$, respectively.}\label{fig:5.1}
\end{figure}
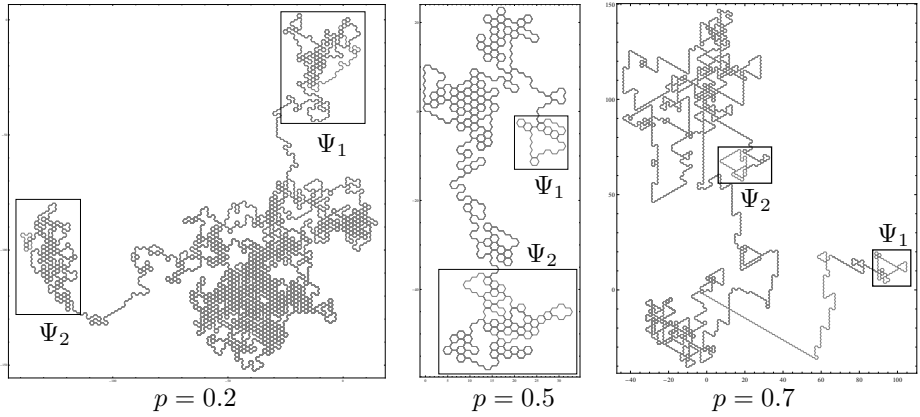

\begin{example}\label{ex:5}
Let $F=\{f_1,f_2\}$, where
$$f_1(p)=\frac{1}{2}\cos(\frac{\pi}{2} p) \ \text{and} \ f_2(p)=1-\frac{1}{2}\cos(\frac{\pi}{2} p).$$
As can be seen in figure \ref{fig:1}, each lattice site $\mathbf{h}\in\mathbb{H}$ is at the end of exactly one horizontal lattice bond. We let $\mathbb{H}^-$ denote the collection of lattice sites $\mathbf{h}\in\mathbb{H}$, that are on the left hand side of a horizontal bond and $\mathbb{H}^+$, those lattice sites $\mathbf{h}\in\mathbb{H}$ that are on the right hand side of a horizontal bond, respectively. If
$$\mathbb{P}[C(\mathbf{h})=1]=
\begin{cases}
f_1(p) \ \text{if} \ \mathbf{h}\in\mathbb{H}^+\\
f_2(p) \ \text{if} \ \mathbf{h}\in\mathbb{H}^-
\end{cases}$$
then, according to corollary \ref{cor:4}, the particle in the $(H;F(p))$ model will be periodic with probability 1, for each $p\in[0,1)$. As an illustration of this, a realization of the particle's trajectory in the $(H;F(p))$ model is shown for $p=0.2$, $0.5$, and $0.7$, in figure \ref{fig:5.1} (cf. figure \ref{fig:3.0}).

Importantly, $f_1(p)\neq f_2(p)$ for each $p\in(0,1]$, so that away from $p=0$, the model's initial orientation of scatterers is not identically distributed. Despite this, the particle's motion in the $(H;F(p))$ model will still be periodic, with probability 1, for $p\in(0,1)$ , which follows from corollary \ref{cor:4}.
\end{example}

\section{Admissible Configurations and Local Correlations}\label{sec:7}
In sections \ref{sec:2}--\ref{sec:5}, we considered the $(H;p)$ model, in which the initial orientation of scatterers was both independent and identically distributed. In section \ref{sec:6} we considered the $(H;F(p))$ model, in which the initial orientation of scatterers was independent but \emph{not} identically distributed. In this section we consider another variation of the $(H;p)$ model, in which we introduce correlations between the initial orientations of the model's scatterers, so that these are no longer independent of each other.

For this model, which we call the $(H;A(p))$ model, we show that although the initial orientation of the scatterers is still assumed to have the same distribution as in the $(H;p)$ model, i.e. each scatterer is a right scatterer with probability $p$, these correlations lead to an absence of reflecting structures, and consequently to a qualitative change in the particle's dynamics. Hence, the assumption that the orientation of each scatterer is independently chosen, is essential to the particle's dynamics, in the $(H;p)$ model.

The initial configuration of scatterers we will consider in the $(H;A(p))$ model will be admissible configurations, which were first introduced in \cite{Webb14}. An admissible configuration $C$ is a particular type of configuration on the honeycomb lattice $H$, in which each hexagon $\mathcal{H}$ of the lattice has a particular orientation of left and right scatterers. To define an admissible configuration we note the following.

\begin{figure}
    \begin{overpic}[scale=.26]{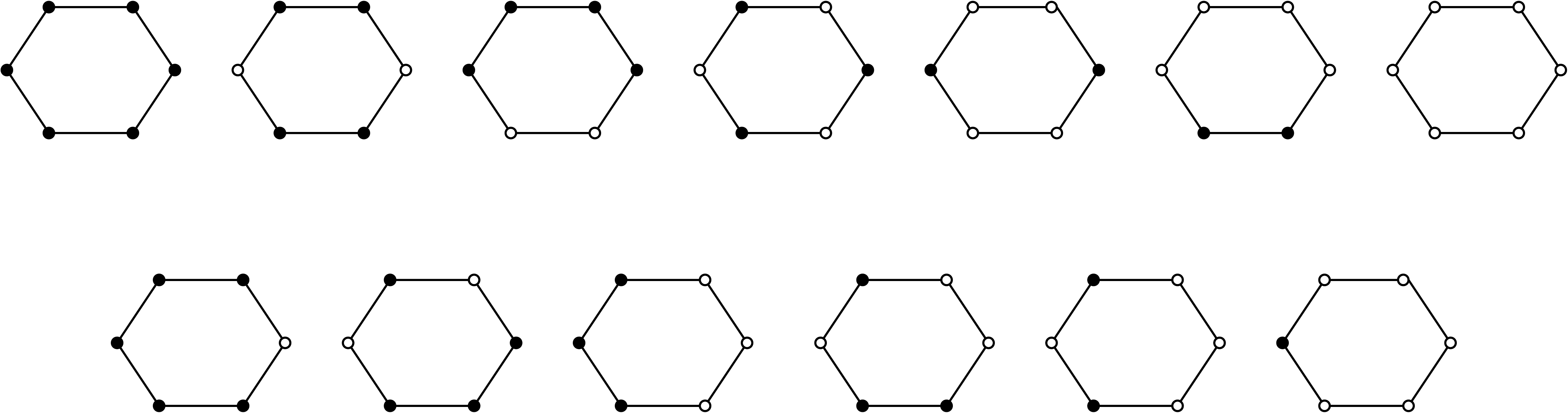}
    \put(26,14.5){\emph{Admissible Hexagonal Configurations}}

    \put(4,21){(1)}
    \put(18.5,21){(2)}
    \put(33.1,21){(3)}
    \put(47.75,21){(4)}
    \put(62.75,21){(5)}
    \put(77.5,21){(6)}
    \put(92.25,21){(7)}

    \put(22,-3){\emph{Non-Admissible Hexagonal Configurations}}

    \put(11,3.5){(8)}
    \put(25.75,3.5){(9)}
    \put(39.75,3.5){(10)}
    \put(55,3.5){(11)}
    \put(69.75,3.5){(12)}
    \put(84.5,3.5){(13)}

    \end{overpic}
\vspace{0.15in}
\caption{The seven admissible hexagonal configurations and the six non-admissible hexagonal configurations are shown. Open and filled circles represent right and left rotators, respectively.}\label{fig:9}
\end{figure}
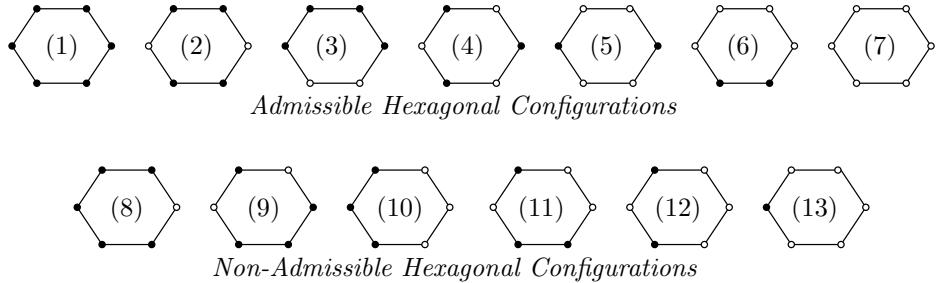

Since there are six lattice sites on any single hexagon of $H$, there are $2^6=64$ possible configurations of left and right scatterers on each hexagon. In fact, up to rotation and reflection there are only thirteen different possible configurations, which are labeled (1)--(13) in figure \ref{fig:9}. These can be separated into two distinct classes (1)--(7) and (8)--(13), which are referred to as \emph{admissible} and \emph{nonadmissible hexagonal configurations}, respectively, as shown in figure \ref{fig:9}.

This distinction between admissible and nonadmissible configurations on a single hexagon can be extended to a configuration $C$ of the entire honeycomb lattice in the following way (cf. \cite{Webb14}).

\begin{definition}\label{def:admiss}
Let $C$ be a configuration of scatterers on the hexagonal lattice. We say that the configuration $C$ is \emph{admissible} if the restriction of $C$ to each hexagon of $H$ has an admissible hexagonal configuration (cf. figure \ref{fig:9}). Otherwise, we say $C$ is a \emph{nonadmissible} configuration.
\end{definition}

Our interest in admissible configurations is due to the following result, found in \cite{Webb14}. It states that, if the $(H,I)$ model has an admissible initial configuration then, between returns to its initial position, the particle's trajectory is a self-avoiding walk.

\begin{theorem}\label{lem:1}\textbf{(Self-Avoiding Motion \cite{Webb14})}
Suppose $(H,I)$ has the initial condition $I=(\mathbf{r},\mathbf{v},C)$ where $C$ is admissible.\\
(a) If the particle has the sequence of times $\{\tau_i\}_{i\in\mathcal{I}}$ at which $\mathbf{r}(\tau_i)=\mathbf{r}$ then each $\gamma_i=T[\tau_{i-1},\tau_i]$ is a self-avoiding cycle based at $\mathbf{r}$ for each $i\in\mathcal{I}$.\\
(b) If $\tau$ is the last time at which $\mathbf{r}(\tau)=\mathbf{r}$, then $\{\mathbf{r}(t)\}_{t\geq \tau}$ is a sequence of distinct positions.
\end{theorem}

Based on theorem \ref{lem:1}, the particle's motion in $(H,I)$ is affected by whether the model's initial configuration is admissible or not. In the $(H;p)$ model, we will see that, with probability 1, the model's initial configuration is nonadmissible for $p\in(0,1)$ but is admissible if $p=0,1$ (see proposition \ref{prop:5}). Therefore, we do not expect to see any self-avoiding behavior in the $(H;p)$ model if $p\in(0,1)$, but are guaranteed this type of behavior if $p=0,1$.

From a physical point of view, a configuration must have a certain degree of \emph{order} to be admissible. Since a randomly generated configuration, in which each orientation is chosen independently, will not be \emph{ordered}, such configurations will not be admissible, e.g. the initial configuration in $(H;p)$ for $p\in(0,1)$. On the other hand, the much more ordered configurations of all left or all right scatterers, respectively, are admissible, e.g. the initial configurations in $(H;p)$ for $p=0,1$. This dichotomy regarding which $p$-values lead to admissible initial configurations and which lead to non-admissible initial configurations in the $(H;p)$ model, is stated as the following proposition.

\begin{proposition}\label{prop:5}
For $p\in[0,1]$, let $\mathbb{P}[C_p=A]$ be the probability that the initial configuration in the $(H;p)$ model is admissible. Then\\
(a) $\mathbb{P}[C_p=A]=0$ for each $p\in(0,1)$; and\\
(b) $\mathbb{P}[C_p=A]=1$ for $p=0,1$.
\end{proposition}

To prove this proposition, we first calculate the probability that any given hexagon of the lattice will begin with an admissible hexagonal configuration. What we show is that, this probability is not equal to 1 if $p\in(0,1)$ and is 1 if $p=0,1$. The idea is that, if any single hexagon has a probability less than 1 of having an admissible configuration, the probability is zero that every hexagon of $\mathbb{H}$ will have an admissible configuration. This summarizes the method used in the following proof of proposition \ref{prop:5}.

\begin{proof}
For $p\in[0,1]$, let $\mathbb{P}[C_p^i(\mathcal{H})]$ be the probability that a hexagon $\mathcal{H}\subset H$ in $(H;p)$ has an initial configuration $(i)$, as shown in figure \ref{fig:9}, up to rotation and reflection, for $1\leq i\leq 7$. Then one can compute that
\[
\mathbb{P}[C^1_p(\mathcal{H})]=(1-p)^6, \ \mathbb{P}[C_p^2(\mathcal{H})]=3p^2(1-p)^4, \ \mathbb{P}[C_p^3(\mathcal{H})]=6p^2(1-p)^4,
\]
\[
\mathbb{P}[C_p^4(\mathcal{H})]=(1-p)^6, \ \mathbb{P}[C_p^5(\mathcal{H})]=3p^2(1-p)^4, \ \mathbb{P}[C_p^6(\mathcal{H})]=6p^2(1-p)^4,
\]
\[
\text{and} \ \ \mathbb{P}[C_p^7(\mathcal{H})]=p^6.
\]
The probability $\mathbb{P}[C_p(\mathcal{H})=A]$, that a hexagon $\mathcal{H}$ initially has an admissible configuration, is then the sum
\[
\mathbb{P}[C_p(\mathcal{H})=A]=\sum_{i=1}^7\mathbb{P}[C_p^i(\mathcal{H})]=1-6p+24p^2-54p^3+72p^4-54p^5+18p^6.
\]
By minimizing this function over all $p\in[0,1]$, it follows that $\mathbb{P}[C_p(\mathcal{H})=A]<1$ for any $p\in(0,1)$. Since there are infinitely many disjoint hexagons on $H$, the Borel-Cantelli lemma (see \cite{Feller86}, page 201) then implies that, $\mathbb{P}[C_p=A]=0$ for each $p\in(0,1)$. Hence, part (a) of proposition \ref{prop:5} holds. Furthermore, since $\mathbb{P}[C_p=A]=1$ for $p=0,1$, then part $(b)$ holds as well.
\end{proof}

Proposition \ref{prop:5} states that those initial configurations, which are randomly generated for $p\in(0,1)$, are nonadmissible, whereas the initial configurations, which are deterministically generated for $p=0,1$, are admissible. Moreover, if the particle moves on the lattice $\mathbb{H}$, which has an admissible configuration, the particle will never encounter a reflecting structure (cf. corollary \ref{cor:2}). Thus, a particle can only encounter a reflector, if the lattice configuration is initially nonadmissible. This is summarized in the following table.

\begin{table}[ht]

\centering

\begin{tabular}{| c | c | c | c |}
\hline
   $p$-values&(I) \ configuration type& (II) \ dynamics& (III) \ reflectors\\
\hline
\hline
   $p\in(0,1)$&probabilistic $\&$ nonadmissible&periodic&yes\\
\hline
    $p=0,1$&deterministic $\&$ admissible&unbounded&no\\
\hline
\end{tabular}
\vspace{0.15in}
\caption{The table shows the differences in the $(H;p)$ model's $(I)$ type of initial configuration, $(II)$ dynamics, and $(III)$ when reflecting structures can or cannot be created by the particle in this model, for different $p$-values, respectively.}
\label{table:0}
\end{table}

Part (I) of table \ref{table:0} seems to suggest that, by generating an initial configuration using some probabilistic rule, the result will always be a nonadmissible configuration. In fact, if we introduce local correlations between the $(H;p)$ model's initial orientations, we can probabilistically generate admissible configurations. Although these configurations will be disordered and random, the fact that they are admissible will mean, via theorem \ref{lem:1}, that the particle's motion through the lattice will be a sequence of self-avoiding cycles. In order to describe the admissible, but randomly generated, configurations, which we will consider in this section, we first note the following.

\begin{figure}
    \begin{overpic}[height=1.5in, width=3.5in]{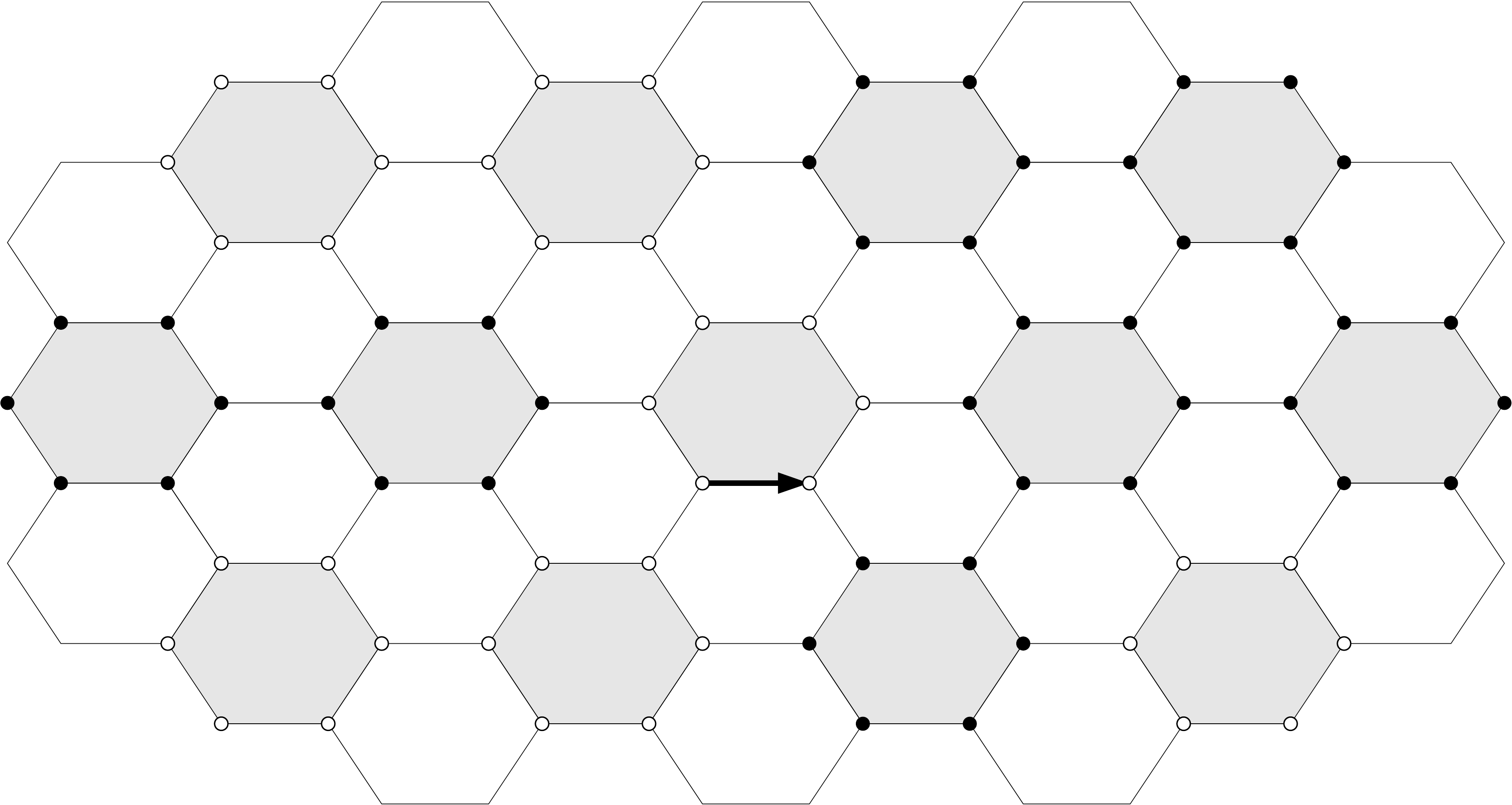}
    \put(43,16){$\mathbf{r}$}
    \put(49,18){$\mathbf{v}$}
    \end{overpic}
\caption{The shaded hexagons, shown above, comprise the set $\mathcal{S}\subset H$ of hexagons, used to generate the random admissible configurations in the $(H;A(p))$ model.}\label{fig:10}
\end{figure}

The honeycomb lattice $H$ consists of an infinite number hexagons, whose sides and vertices make up the lattice's bonds and sites, respectively. Let $\mathcal{S}$ denote the set of hexagons on the honeycomb lattice, which are \emph{shaded} in figure \ref{fig:10}. As can be seen in this figure, each lattice site of $H$ belongs to exactly one hexagon $\mathcal{H}\in\mathcal{S}$. If we specify the orientation of each scatterer on the hexagon $\mathcal{H}$ for all $\mathcal{H}\in\mathcal{S}$, this will generate a configuration of scatterers on the entire honeycomb lattice $H$. The following model can then be defined.

Let $(H;A(p))$ denote the flipping rotator model, in which the particle has the initial position $\mathbf{r}=(0,0)$ and velocity $\mathbf{v}=(1,0)$. The model's initial configuration $C$ is the random initial configuration where, for each $p\in[0,1]$ and $\mathcal{H}\in\mathcal{S}$, all scatterers of $\mathcal{H}$ are oriented to the right with probability $p$ and to the left with probability $1-p$, respectively. Furthermore, we assume that the initial orientation of the scatterers on any single hexagon of $\mathcal{S}$ does not influence the initial orientation of the scatterers on any other hexagon of $\mathcal{S}$, i.e. they are independently generated.

Under these assumptions, the initial orientation of scatterers in the $(H;A(p))$ model are identically distributed, since each scatterer has the probability $p$ of being a right scatterer and probability $1-p$ of being a left scatterer. These orientations, however, are not independently chosen but are \emph{correlated}, since each scatterer on any hexagon of $\mathcal{S}$ has the same initial orientation in this model.

A specific consequence of the correlations found in the $(H;A(p))$ model is that, any realization of the initial configuration in this model will automatically be an admissible configuration (cf. figure \ref{fig:9}). Theorem \ref{lem:1} then implies that the particle in the $(H;A(p))$ model will have a self-avoiding motion away from its initial position. This self-avoiding behavior is described in the following corollary of theorem \ref{lem:1}, which also states that reflecting structures cannot be formed in the $(H;A(p))$, for any value of $p$.

\begin{corollary}\label{cor:5}
For any $p\in[0,1]$, there will be, with probability 1, a sequence of times $\{\tau_i\}_{i\in\mathcal{I}}$ in the $(H;A(p))$ model, such that $\mathbf{r}(\tau_i)=\mathbf{r}$ and the following hold:\\
(a) Each $\gamma_i=T[\tau_{i-1},\tau_i]$ is a self-avoiding cycle, based at $\mathbf{r}$ for each $i\in\mathcal{I}$.\\
(b) If $\tau$ is the last time $\mathbf{r}(\tau)=\mathbf{r}$, then $\{\mathbf{r}(t)\}_{t\geq \tau}$ is a sequence of distinct positions.\\
(c) The particle never encounters a reflecting structure at any time $t\geq0$.
\end{corollary}

The proof of this result is based on the fact, that the initial configuration in the $(H;A(p))$ model will always be admissible. Hence, the particle in this model will have a self-avoiding motion (see theorem \ref{lem:1}). Because of this, the particle will never encounter a reflecting structure since, in order to form a reflector, the particle has to cross its own path away from its initial position. This is impossible, if the particle's motion is self-avoiding. This summarizes the following proof of corollary \ref{cor:5}.

\begin{proof}
Since any realization of the initial configuration in the $(H;A(p))$ model is admissible for $p\in[0,1]$, (a) and (b) follow directly from theorem \ref{lem:1}, while (c) follows using the same argument as is given in the proof of corollary \ref{cor:2}.
\end{proof}

\begin{figure}
\begin{center}
\begin{tabular}{cc}
    \begin{overpic}[scale=.18]{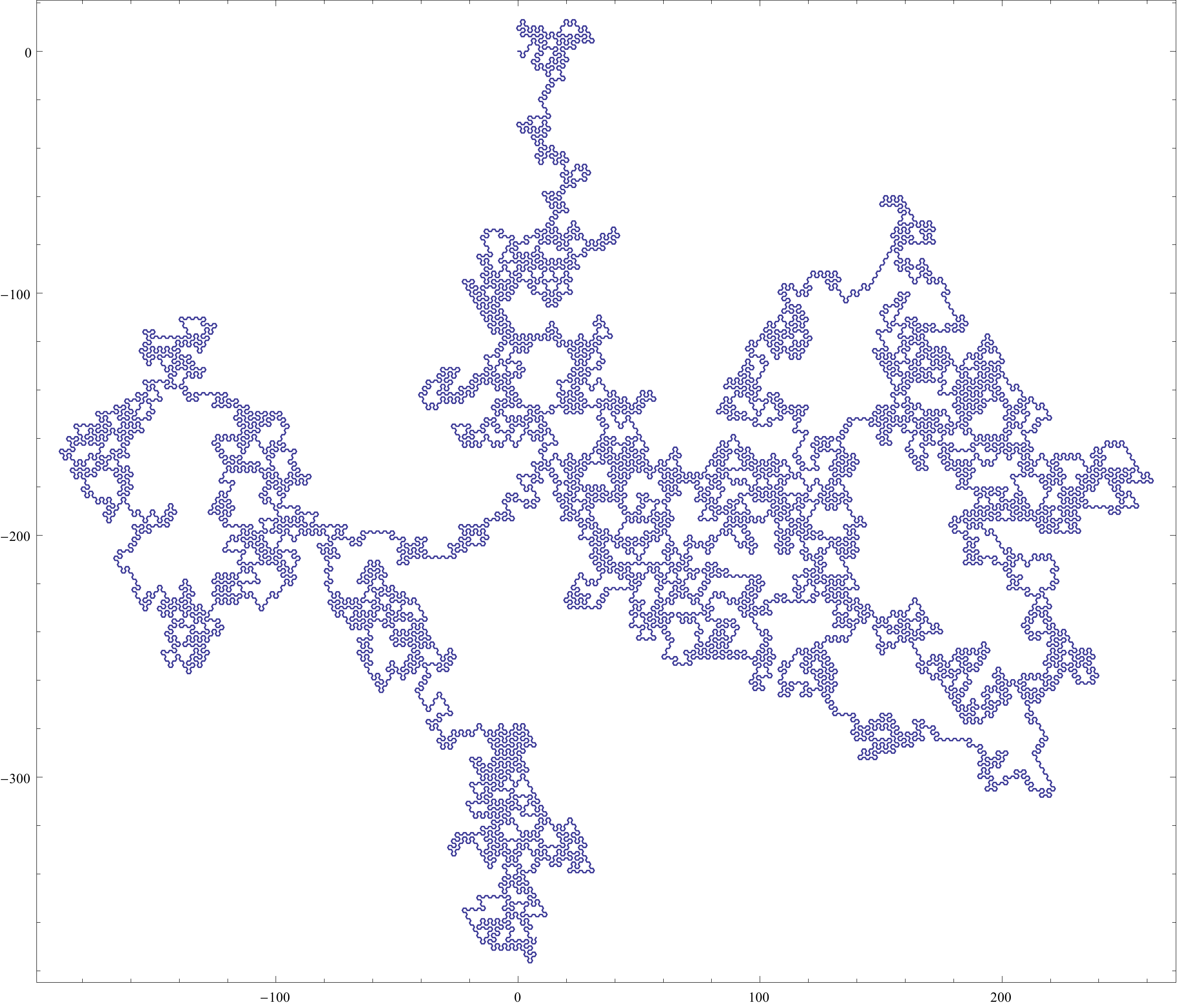}
    \put(48,-7){$(a)$}
    \end{overpic} &
    \begin{overpic}[scale=.75]{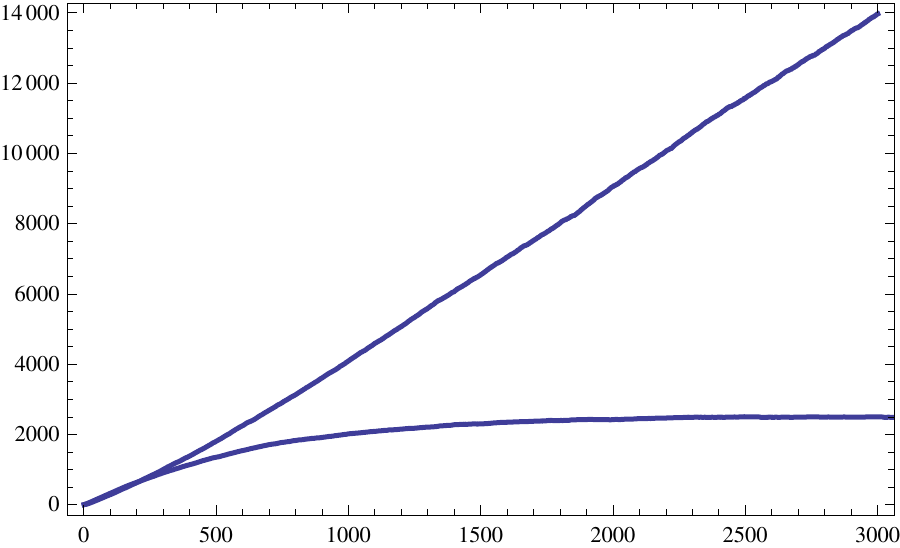}
    \put(65,17){$\triangle(t)$}
    \put(45,37){$\triangle_A(t)$}
    \put(48,-5){$(b)$}
    \end{overpic}
\end{tabular}
\end{center}
\caption{In (a) a realization of the particle's trajectory in the $(H;A(1/2))$ model is shown for $t\leq 20000$. In (b) the difference between the particle's mean square displacement $\triangle(t)$ and $\triangle_A(t)$ in the $(H;1/2)$ and $(H;A(1/2))$ models are shown, respectively, for $t\leq 3000$.}\label{fig:7.1}
\end{figure}

As an illustration of corollary \ref{cor:5}, a realization of the particle's trajectory in the $(H;A(1/2))$ model is shown in figure \ref{fig:7.1}(a) for $t\leq 20000$. Since the particle does not return to the origin by the time $t=20000$ in this realization, the particle's entire trajectory, up to this point in time, is a self-avoiding walk, as guaranteed by corollary \ref{cor:5}.

The major difference between the $(H;p)$ model and the $(H;A(p))$ model is that, in the latter, the scatterers on each hexagon $\mathcal{H}\in\mathcal{S}$ are \emph{not} independently chosen to be either right or left scatterers, but are chosen to be either all left scatterers or all right scatterers. Despite this difference, the initial configurations of scatterers in both these models are identically distributed and, in fact, these distributions are the same, i.e. in both models $\mathbb{P}[C(\mathbf{h})=1]=p$ for any $\mathbf{h}\in\mathbb{H}$. Additionally, we have that $(H;p)=(H;A(p))$ for $p=0,1$. Therefore, the only difference between these two models are the local correlations found in the initial scattering configuration of the $(H;A(p))$ model.

What we observe in the $(H;A(p))$ model, via numerical experiments, is that the absence of reflecting structures leads to a lack of periodicity in the particle's motion. This can be seen in figure \ref{fig:7.1}(b), where the particles's mean square displacement is plotted for $p=1/2$ in both the $(H;A(p))$ and the $(H;p)$ models. The particle's \emph{mean square displacement} is given by:
\begin{align*}
\triangle(t)\equiv&\langle|\mathbf{r}(t)|^2\rangle \ \text{for} \ t\geq 0 \ \text{in the} \ (H;p) \ \text{model}; \ \text{and}\\
\triangle_A(t)\equiv&\langle|\mathbf{r}_A(t)|^2\rangle \ \text{for} \ t\geq 0 \ \text{in the} \ (H;A(p)) \ \text{model,}
\end{align*}
where $\mathbf{r}(t)$ denotes the particle's position in the $(H;p)$ model and $\mathbf{r}_A(t)$ denotes the particle's position in the $(H;A(p))$ model. The average $\langle\cdot\rangle$ is taken over all realizations of the random initial condition $I=(\mathbf{r},\mathbf{v},C)$, in each respective model. The norm $|\cdot|$ is the standard Euclidean distance in the plane.

In figure \ref{fig:7.1}(b), we see that the particle's mean square displacement $\triangle(t)$ in the $(H;1/2)$ model asymptotically approaches the value $\mathcal{D}=2000$ as a consequence of the particle's periodic motion. Physically, the value $\mathcal{D}$ is the average distance the particle is away from its initial position during one period of its motion, in the $(H;1/2)$ model. In contrast, the particle's mean square displacement $\triangle_A(t)$ in the $(H;A(1/2))$ model, can be numerically approximated by the function $\triangle_A(t)\approx t^{1.18}$, which suggests that the particle in this model has an unbounded trajectory.

Thus, by introducing local correlations into the initial configuration of the $(H;p)$ model, it is possible to drastically change the particle's motion from being periodic to what numerically appears to be a non-periodic super-diffusive motion for $p=1/2$ (see figure \ref{fig:7.1}(b)). This may seem somewhat surprising, since the correlations found in the $(H;A(p))$ model's initial configuration are \emph{local}, in that only scatterers on the same hexagon can be correlated. Yet, these correlations are sufficient to change the \emph{global} properties of the particle's dynamics.

\section{Conclusion}\label{sec:8}
In this paper we have considered the deterministic motion of a particle moving on the honeycomb lattice $H$, whose sites are randomly occupied by left and right flipping rotators. When each individual scatterer is initially oriented to the right, independent of the other scatterers with probability $p\in(0,1)$, we show that the particle will have a periodic trajectory, with probability 1. In contrast, if $p=0,1$ the particle's motion is a sequence of self-avoiding cycles between returns to its initial position, which numerical simulations suggest to be unbounded and, therefore, non-periodic.

When $p\in(0,1)$, the particle can, as it moves through the lattice, create a number of structures. These structures, which we refer to as \emph{reflecting structures} (see section \ref{sec:2}), cause the particle to reverse its entire trajectory back through every one of the lattice sites it has previously visited. Importantly, two of these reflectors can trap the particle, limiting its motion to a finite subset of the lattice and cause the particle to have a periodic trajectory (see theorem \ref{thm:1}).

These reflecting structures and their role in causing periodic dynamics, have been previously observed in this and other LLG models \cite{Bunimovich93,Wang95.1}. In this paper though we describe, how the particle not only \emph{creates} these structures, but how the particle can also \emph{transform} and \emph{annihilate} them. By studying the interplay of these three processes of creation, transformation, and annihilation, we are also lead to introduce the concepts of semi-reflecting structures as well as transforms of reflecting structures (see section \ref{sec:3}). The processes of reflector creation, transformation, and annihilation, as well as the notions of semi-reflecting structures and reflector transforms, are each important for the particle's dynamics and are used to show, that the particle's motion will be periodic for $p\in(0,1)$, with probability 1 (see section \ref{sec:4}).

In contrast, when $p=0,1$, as was considered in \cite{Webb14}, the particle, between returns to its initial position, has a self-avoiding motion. Since this self-avoiding motion also appears to be unbounded, the difference between the particle's periodic self-limiting motion, which occurs for $p\in(0,1)$, and its self-avoiding motion, which happens for $p=0,1$, is quite striking. Hence, as $p$ approaches either $0$ or $1$, what we observe is that the particle undergoes a \emph{discontinuous} dynamical transition from a periodic, self-limiting motion to a non-periodic self-avoiding mode of motion.

By combining the results in this paper, regarding the particle's self-limiting motion, with the theory previously developed in \cite{Webb14}, which describes the particle's self-avoiding behavior, we are able to qualitatively describe the particle's transition between these two types of motion. For $p$ close to 0 or 1, we find that, while the particle's transient, or short term, dynamics consists of a finite number of self-avoiding cycles, its long term dynamics is ultimately periodic. As $p$ approaches 0 and 1, i.e. when the initial orientations of scatterers become more homogenous, we observe that the particle's period becomes increasingly longer, so that, in the limit as $p$ approaches either 0 or 1, the particle's motion is no longer bounded, but is entirely composed of an infinite sequence self-avoiding cycles.

Furthermore, we have shown that for $p=0,1$, it is not possible for the particle to form reflecting structures, suggesting that an absence of reflecting structures leads to non-periodic behavior. Stated another way, as the medium through which the particle moves, becomes increasingly homogenous, the particle becomes less able to build reflectors until at $p=0,1$, where the lattice is completely homogeneous, these structures can no longer exist.

To better understand why random initial configurations allow the particle to form reflecting structures, we also consider in this paper a generalization of our original $(H;p)$ model. In that model, which we refer to as $(H;F(p))$, the initial configurations of the scatterers are \emph{not} identically distributed, but are still independently chosen. We find that even without being identically distributed, the particle's trajectory will still be periodic (see section \ref{sec:6}). In contrast, if the initial orientations of the scatterers are \emph{not} chosen independently, this can drastically change the particle's dynamics, even if these orientations are identically distributed (see section \ref{sec:7}). Hence, the particle's motion in the LLG we consider in this paper, although deterministic, depends on the particular statistical properties of the model's initial configuration.

The findings in this paper also lead to a number of open questions. One is, whether admissible configurations, including the random admissible configurations introduced in this paper, \emph{always} lead to non-periodic and therefore unbounded motion. Numerically, we observe this, but whether this can be proved mathematically, is unknown. This is related to a second open question, regarding the geometric nature of the self-avoiding cycles found in those flipping rotator models with an admissible initial configuration. In each case, these cycles appear to have a fractal-like structure, but it is unknown to what extent this is, or is not, the case. Indeed, an understanding of the geometry of these cycles and their dependence on the system's initial configuration could be important in a number of applications, since self-avoiding walks are used to model the growth of crystals, polymers, etc. \cite{Amit83,Madras13,Bous92,Guttmann12}.

Additionally, there are questions regarding the reflecting structures, that are found in the models considered here. For instance, one of these questions is whether, reflecting structures are the only mechanism that can cause periodic motion. Indeed, if this could be shown to be the case, then we have an affirmative answer to the open question, whether or not admissible configurations \emph{always} lead to unbounded trajectories, via corollary \ref{cor:2} (see section \ref{sec:5}).

As we plan to show in later papers, the notions of reflector creation, transformation, and annihilation, are also relevant for other two-dimensional LLG models, that have yet to be considered in the literature.

\end{spacing}

\end{document}